\documentclass[amsmath,amssymb,aps,prresearch,twocolumn,longbibliography,superscriptaddress]{revtex4-2}

\usepackage{hyperref}
\usepackage[pdftex,dvipsnames]{xcolor}
\usepackage{bm}
\usepackage{graphicx}
\usepackage{euscript}
\usepackage{lipsum}                     
\usepackage{xargs} 
\usepackage[dvipsnames]{xcolor}
\usepackage[normalem]{ulem} % for \sout (strikethrough)
\usepackage{float}
\usepackage{soul}

\newcommand{\added}[1]{\textcolor{black}{#1}}
\newcommand{\add}[1]{{\textcolor{black}#1}}
\newcommand{\removed}[1]{\if0{#1}\fi}

\newcommand{\soften}[1]{\textcolor{black}{#1}}
\newcommand{\new}[1]{\textcolor{black}{#1}}

\begin{document}

\title{ \soften{Symmetry-Protected Topological Structures in Quantum Many-Body Scar Subspaces}}% Force line breaks with \\

\author{Chihiro Matsui}
 \affiliation{Department of Mathematical Sciences, The University of Tokyo, 3-8-1 Komaba, Meguro-ku, Tokyo 153-8914, Japan}%Lines break automatically or can be forced with \\
\author{Thomas Quella}%
 \affiliation{School of Mathematics and Statistics, The University of Melbourne, Parkville, VIC 3010, Australia}%
\author{Naoto Tsuji}%
\affiliation{Department of Physics, University of Tokyo, Bunkyo-ku, Tokyo 113-8656, Japan}
\affiliation{RIKEN Center for Emergent Matter Science (CEMS), Wako, Saitama 351-0198, Japan}
\affiliation{Trans-scale Quantum Science Institute, University of Tokyo, Bunkyo-ku, Tokyo 113-8656, Japan}

\date{\today}% It is always \today, today,

\begin{abstract}

\soften{
We investigate how symmetry-associated topological properties, originally defined for ground states, can be consistently extended to dynamically isolated subspaces of quantum many-body systems. Focusing on quantum many-body scars (QMBS), we show that a scar subspace is invariant under the relevant symmetries, and it can support well-defined topological properties inherited from an underlying symmetry-protected topological (SPT) ground state, even though individual scar states need not themselves be symmetric. We refer to such structures as SPT-like scar subspaces, emphasizing that the topological properties are defined at the level of the symmetry-invariant subspace rather than on a state-by-state basis.
As a concrete example, we analyze the spin-1 Affleck–Kennedy–Lieb–Tasaki (AKLT) model, where a restricted spectrum-generating algebra (rSGA) generates a tower of QMBS on top of a nontrivial SPT ground state admitting a matrix product state (MPS) representation. We demonstrate that the resulting symmetry-invariant scar subspace inherits well-defined topological characteristics, encoded in bond-space symmetry representations, expectation values of symmetry-twist operators, and long-range string order in the dilute regime. We further show that these features persist in a broader class of models, including deformations and higher-spin generalizations that preserve the relevant symmetries and admit nontrivial SPT ground states with MPS representations.
Our results provide a general framework in which symmetry-associated topological properties, while inherited from the ground state, are consistently realized at the level of symmetry-invariant nonthermal subspaces. This perspective clarifies how topological structures can be meaningfully defined beyond individual eigenstates.
}

\end{abstract}

\maketitle

%\tableofcontents

\section{Introduction}

Topological phases of matter provide a unifying framework for understanding quantum many-body systems that extend beyond Landau's symmetry-breaking paradigm~\cite{Wen1990, WenBook2004, Bhardwaj:2024PhRvL.133p1601B}. 
Among these, symmetry-protected topological (SPT) phases~\cite{GuWen2009, HasanKane2010, QiZhang2011, Chen2012Science, Chen2013} occupy a particularly fundamental role. 
An SPT phase is characterized by a unique, gapped ground state that remains invariant under on-site or space-time symmetries, and by discrete topological invariants that classify distinct phases~\cite{ChenGuWen2011,Chen2013,Pollmann2010,Pollmann2012,Schuch2011}. 
Different SPT phases cannot be adiabatically connected without closing the energy gap, as long as the protecting symmetry is preserved~\cite{ChenGuWen2011,Chen2013,Pollmann2010,Pollmann2012,Schuch2011}.
One-dimensional quantum spin chain models serve as a representative example of SPT phases~\cite{Pollmann2010, Chen2011, Pollmann2012, Schuch2011,Duivenvoorden:2012arXiv1206.2462D}. Ground states of such phases are known to exhibit low entanglement and admit an efficient (and often accurate) representation in terms of matrix product states (MPS)~\cite{Hastings:2007JSMTE..08...24H,DeRoeck:2025arXiv251114699D}.

A natural motivation \soften{for extending the topological properties associated with symmetries beyond ground states} comes from the remarkable structure of quantum many-body scars (QMBS)~\cite{Bernien2017, Shiraishi2017, Turner2018, Turner2018b, Serbyn2021, Moudgalya2022, Chandran2023}, which are exceptional nonthermal eigenstates of an otherwise nonintegrable Hamiltonian.  
Despite being highly excited, QMBS exhibit several features usually associated with ground states: 
they retain long-lived oscillations—i.e., persistent memory of the initial state~\cite{Bernien2017,Turner2018,Turner2018b,Serbyn2021,Iadecola:2019PhRvB.100r4312I}—and have entanglement entropy far below those of generic excited states
~\cite{Moudgalya2018,Chattopadhyay:2020PhRvB.101q4308C,Mark2020,Serbyn2021}.

Due to these quasi–ground-state properties, it is natural to \added{ask whether} phenomena traditionally restricted to ground states—including symmetry-protected topological characteristics—could emerge within the corresponding scar subspaces. 
\add{In fact, there exist constructions of scar-like highly excited states that exhibit nontrivial topological properties in the sense of SPT phases, obtained by embedding or perturbatively deforming Hamiltonians with topologically nontrivial ground states so as to promote such ground states into the middle of the spectrum~\added{\cite{Ok2019}}, or by constructing states that approximately satisfy a restricted spectrum-generating algebra (rSGA) on top of a reference state exhibiting nontrivial SPT properties~\cite{Jeyaretnam2021}. 
In these constructions, however, the protecting symmetry was not explicitly used to characterize their SPT properties, and it therefore remains unclear whether such topological properties merely reflect the nontrivial SPT nature of the ground state, or whether the QMBSs themselves admit a corresponding SPT characterization based on cohomological classification.}

\add{The focus of this work is to demonstrate that SPT-like structures associated with the protecting symmetry can be identified at the level of dynamically isolated scar subspaces themselves, using the conventional cohomological classification of ground-state SPT phases.}
This \added{perspective} becomes particularly compelling in scar subspaces stabilized by a restricted spectrum-generating algebra (rSGA)~\cite{Moudgalya:2020PhRvB.102h5140M}. 
The rSGA dynamically isolates a tower of scar states from the surrounding thermal continuum and enforces an equally spaced spectrum within the tower, thereby making the scar subspace behave as an effectively protected sector embedded inside a nonintegrable spectrum.  In many constructions, the reference state of the scar tower is either a simple product state or a low-bond-dimension MPS~\cite{Moudgalya2018,Chattopadhyay:2020PhRvB.101q4308C,Lin2020,Mark2020}\added{, including cases where it corresponds to a nontrivial SPT ground state~\cite{Moudgalya:2018PhRvB..98w5156M}}. Moreover, individual scar states often admit a quasiparticle description~\cite{Chandran2023}. Together, these properties \soften{provide a natural setting in which} MPS-based characterization of \soften{symmetry-associated topological properties can be extended} to the entire tower. \added{In particular, when the reference state is a nontrivial SPT ground state, the resulting symmetry-invariant scar subspace is expected to retain its topological signatures at the level of the entire subspace, while remaining accessible from experimentally relevant initial states~\cite{Zhao2021OrthogonalScars,Gustafson2023PrepScars,Moudgalya2022}. }

\soften{In this work, we show that the symmetry-invariant scar subspace as a whole can exhibit topological properties inherited from the SPT ground state, either directly or in a modified form. We further show that these properties become well defined at the level of the scar subspace, where the protecting symmetries act consistently. We refer to such symmetry-invariant scar subspaces with well-defined topological properties as \emph{SPT-like scar subspaces}.}

As a concrete example, we focus on the spin-$1$ Affleck–Kennedy–Lieb–Tasaki (AKLT) model~\cite{Affleck1987,Affleck1988}, where a nontrivial SPT ground state coexists with an exact rSGA structure and thus provides an ideal platform for testing how topological properties manifest within the scar subspace.  
We analytically show that the resulting scar subspace exhibits well-defined \added{symmetry-associated} topological \added{properties}, confirmed by its bond-space symmetry representations, \added{expectation values of symmetry-twist operators}, and long-range string order. We further demonstrate that \new{these features persist across a broader class of models, including deformed AKLT models and higher-spin generalizations that preserve the associated symmetries, as well as in symmetry-invariant scar subspaces constructed on nontrivial SPT ground states}.

The remaining parts of this paper are organized as follows. 
In Sec.~\ref{sec:MPS-SPT}, we review the MPS-based \added{characterization of symmetry-associated topological properties} and show how the same \added{MPS}-based framework can be extended to \added{general symmetry-invariant MPS}.
In Sec.~\ref{sec:SPT-scar}, we formulate the notion of an SPT\added{-like} scar subspace by demonstrating that \added{symmetry-associated} topological properties become well defined when the quasiparticle excitations admit an MPS representation and the subspace is invariant under the protecting symmetries.
In Sec.~\ref{sec:AKLT-scar}, we illustrate these ideas in the AKLT model by
analyzing the bond-space symmetry representations, \added{expectation values of symmetry-twist operators}, and the \added{analytically established} long-range string order \added{in the dilute regime, together with numerical evidence that QMBS exhibit string correlations distinct from those of typical thermal states}.
In Sec.~\ref{sec:robustness}, we \new{show that these topological properties remain well defined under inversion, time-reversal, and on-site $\mathbb{Z}_2 \times \mathbb{Z}_2$ symmetries, for deformations and higher-spin generalizations of the AKLT model. }

\section{Matrix product SPT states}
\label{sec:MPS-SPT}

In this section, we review the MPS-based \added{framework for characterizing topological properties associated with on-site, spatial, or time-reversal symmetry, in terms of} bond-space symmetry representations. We then show how the same framework extends naturally to quasiparticle excitation states that admit an MPS representation. 
The notation and conventions introduced here will be used throughout the paper.

\subsection{Basic definitions}

\added{
We consider translationally invariant matrix product states (MPS) of the form
\begin{align} \label{eq:MPS_homo}
|\psi\rangle = \sum_{m_1,\dots,m_L} 
{\rm tr} (A_{m_1} \cdots A_{m_L})\, |m_1,\dots,m_L\rangle,
\end{align}
where $|m\rangle$ ($m=0,\dots,2s$) denotes a local basis of the physical Hilbert space, and $A_m$ are $D\times D$ matrices acting on a bond space $\mathcal{V}_a$ of dimension $D$. For later convenience, we also introduce the vectorized notation
\begin{align}
|\psi\rangle = {\rm tr}_a (\vec{A} \otimes_p \cdots \otimes_p \vec{A}),
\end{align}
where $\vec{A}$ denotes the collection of matrices $\{A_m\}$, $\otimes_p$ is the tensor product in the physical space, and ${\rm tr}_a$ denotes the partial trace over the bond space $\mathcal{V}_a$.
}

\added{
Such an MPS representation is not unique due to gauge freedom: for any invertible matrix $V$, the gauge transformation $\vec{A} \mapsto V \vec{A} V^{-1}$, acting componentwise on $A_m$, leaves the state invariant. This gauge structure plays a central role in characterizing symmetry actions on the bond space.
Injectivity~\cite{PerezGarcia2007} guarantees that an MPS representation is unique up to a gauge transformation. It is defined by the condition that the map
\begin{align}
\mathcal{A}_\ell:\,
X \mapsto \sum_{m_1,\dots,m_\ell} 
{\rm tr}(X A_{m_1}\cdots A_{m_\ell}) |m_1,\dots,m_\ell\rangle
\end{align}
is injective for some finite $\ell$. 
} %end \added

\subsection{Topological properties under symmetry actions}

Topological properties can be extracted directly from the MPS representation whenever the state admits appropriate on-site or space–time symmetry~\cite{Pollmann2010, Schuch2011, Pollmann2012}.
This perspective enables a Hamiltonian-independent characterization of topological features in MPS.

For a translationally-invariant injective MPS with an on-site symmetry $G$, each
element $g\in G$ acts on the local tensors through its one-site physical 
representation $u_g$ as
\begin{align} \label{eq:phys-to-aux}
g:~
\vec{A} \mapsto u_g \vec{A}
= e^{i\theta_g}\, V_g\, \vec{A}\, V_g^{\dagger},
\end{align}
where $V_g$ forms a (possibly projective) representation of $G$. The projective nature is encoded in the composition rule
\begin{align} \label{eq:proj}
V_{g_j} V_{g_k} = \omega(g_j,g_k) V_{g_j g_k}, \qquad g_j,g_k \in G,
\end{align}
where the $U(1)$ factor $\omega(g_j,g_k)$ satisfies the associativity
constraint and defines a 2-cocycle belonging to the cohomology class
$\{\omega(g_j,g_k)\} \in H^2(G,U(1))$~\cite{Pollmann2010,Chen2011,
Chen2013}.
In one-dimensional bosonic systems with an Abelian on-site symmetry,
projective representations can equivalently be characterized by commuting
phase factors
\begin{align}
&V_{g_j} V_{g_k} = e^{i\phi(g_j,g_k)} V_{g_k} V_{g_j}, \\
&e^{i\phi(g_j,g_k)} =
\frac{\omega(g_j,g_k)}{\omega(g_k,g_j)}, \label{eq:commuting_phase}
\end{align}
which for $G=\mathbb{Z}_2\times\mathbb{Z}_2$ reduce to
$e^{i\phi(g_j,g_k)}=\pm1$, corresponding to $H^2(G,U(1))=\mathbb{Z}_2$.

For inversion ($\mathcal{P}$) and time-reversal ($\mathcal{T}$), the induced actions on the MPS tensors take the form 
\begin{align}
&\mathcal{P}:\quad 
\vec{A} \mapsto {^{t_a}\!}\vec{A}
= e^{i\theta_{\mathcal{P}}}\,
V_{\mathcal{P}}\, \vec{A}\, V_{\mathcal{P}}^{\dag}, \\[3pt]
&\mathcal{T}:\quad 
\vec{A} \mapsto u_{\mathcal{T}} \vec{A}^{\,*}
= e^{i\theta_{\mathcal{T}}}\,
V_{\mathcal{T}}\, \vec{A}\, V_{\mathcal{T}}^{\dag}.
\end{align}
Here ${}^{t_a}\!\vec{A}$ denotes the transposition acting on the bond indices, and $u_{\mathcal{T}}$ is the unitary part of the physical time-reversal operator.
As in the on-site case, these symmetries can act linearly or projectively on the bond space.
The corresponding indicators are given by
\begin{align}
&V_{\mathcal{P}} V_{\mathcal{P}}^{*} = \pm \bm{1}, \\
&V_{\mathcal{T}} V_{\mathcal{T}}^{*} = \pm \bm{1},
\label{eq:P_T_projective_condition}
\end{align}
where the minus sign signals a nontrivial projective action of
$\mathcal{P}$ or $\mathcal{T}$~\cite{Pollmann2010,Pollmann2012}. 
Thus, inversion and time-reversal symmetries also fit naturally into the same cohomology-based framework used to \added{characterize topological properties} of on-site symmetric MPS. 

Throughout this paper, we characterize the topological properties of a matrix product representation—either an MPS for states or an MPO for quasiparticle creation operators—by how the protecting symmetry (on-site, inversion, or time-reversal) is represented on the bond space\added{, in a manner consistent with a cohomological characterization of the resulting topological properties.
We refer to these as \textit{symmetry-associated topological properties}, as they are defined through symmetry actions on the bond space. } Linear representations correspond to \textit{trivial} topology, whereas projective representations signal \textit{nontrivial} topology. 
The latter give rise to familiar signatures such as long-range string order and \added{expectation values of symmetry-twist operators that deviate from unity}~\cite{denNijs1989, KennedyTasaki1992, Pollmann:2012PhRvB..86l5441P,Zaletel:2014JSMTE..10..007Z}.

\subsection{Topological properties of quasiparticle excitations}

\added{The characterization of symmetry-associated topological properties in terms of bond-space representations of MPS can be extended to excited states that admit an MPS representation. }
In particular, a wide class of quasiparticle excitations admits finite-bond-dimension MPS descriptions, allowing us to analyze how they transform under the protecting symmetries. 

A quasiparticle creation operator with momentum $k$ takes the form
\begin{align} \label{eq:q-creation}
Q^{\dag}(k) = \sum_{x=1}^L e^{ikx} q_x^{\dag},
\end{align}
which admits a finite-bond MPO representation~\cite{Moudgalya:2018PhRvB..98w5156M}:
\begin{align}
&Q^{\dag}(k) = {}_b\langle \uparrow\!| M(1;k)\cdots M(L;k) |\!\downarrow \rangle_b, \nonumber \\
&M(x;k) = \begin{pmatrix} e^{ik}\bm{1}_x & e^{ik}q_x^{\dag} \\ 0 & \bm{1}_x \end{pmatrix}_b, \label{eq:MPO_local}
\end{align}
where $x$ refers to the $x$th physical site and the subscript $b$ indicates that the boundary vectors and matrices act on the auxiliary space $\mathcal{V}_b=\mathbb{C}^2$.
In what follows, we focus on quasiparticle excitations at momentum $k=\pi$, for which the creation operator $Q^{\dag}(k)$ will simply be denoted $Q^{\dag}$. At this special momentum, different quasiparticles do not interact—up to their hard-core constraint—so that an $N$-particle state
\begin{align} \label{eq:q-state}
|\Psi_N\rangle = (Q^\dag)^N |\Psi_0\rangle
\end{align}
admits an MPS representation whose bond dimension does not exceed $2^N D$\added{, where the factor $2^N$ arises from the bond dimension of the MPO representation of the creation operator}~\cite{Vidal2003,PerezGarcia2007,Schuch2008}.

If the quasiparticle creation operator $Q^\dag$ respects a protecting symmetry $\mathcal{S}$—\textit{on-site} ($g \in G$), \textit{inversion} ($\mathcal{P}$), or \textit{time-reversal} ($\mathcal{T}$)—in the sense that
\begin{align} \label{eq:q_sym}
\mathcal{S}:~ Q^\dag \mapsto e^{i\rho_{\mathcal{S}}} Q^\dag,
\end{align}
then the resulting quasiparticle state $|\Psi_N\rangle$ is also symmetric under the same symmetry $\mathcal{S}$. Its topological properties can thus be readily defined from the matrix-product representation. 

The induced bond-space transformations for both the underlying ground-state MPS and the quasiparticle-creation MPO take the form 
\begin{align}
&\mathcal{S}:~ \vec{A} \mapsto e^{i\theta_{\mathcal{S}}} 
    V_{\mathcal{S}} \vec{A} V_{\mathcal{S}}^{\dag}, \\
&\mathcal{S}:~ M \mapsto e^{i\theta'_{\mathcal{S}}} 
    V'_{\mathcal{S}} M V'^{\dag}_{\mathcal{S}}. 
\end{align}
Then the $N$-quasiparticle MPS transforms as
\begin{align} \label{eq:MPO_transf}
\mathcal{S} :~
\mathbb{M}\vec{A}
    \mapsto 
e^{i\widetilde{\theta}_{\mathcal{S}}}
(\mathbb{V}'_{\mathcal{S}} \otimes_a V_{\mathcal{S}})
    \mathbb{M}\vec{A}
(\mathbb{V}'^{\dag}_{\mathcal{S}} \otimes_a V^{\dag}_{\mathcal{S}}),
\end{align}
where $\widetilde{\theta}_{\mathcal{S}} = 
\theta_{\mathcal{S}} + N \theta'_{\mathcal{S}}$ is the total phase, and $\mathbb{V}'_{\mathcal{S}} = (V'_{\mathcal{S}})^{\otimes_b N}$ acts on the auxiliary MPO spaces.

As in the ground-state case, the linear or projective nature of $V'_{\mathcal{S}}$ determines the symmetry-associated topological properties of the $N$-quasiparticle states $(Q^\dag)^N|\Psi_0\rangle$. A nontrivial SPT ground state combined with a linear $V'_{\mathcal{S}}$ yields nontrivial topology for all quasiparticle numbers $N$, directly inheriting the topological properties of the ground-state MPS. 
In contrast, a projective $V'_{\mathcal{S}}$ can produce either trivial or nontrivial topology depending on the value of $N$ (Table~\ref{tab:topo_prop}).

Based on this \added{observation}, we refer to the scar subspace as \textit{topologically homogeneous} when its topological properties remain identical for all quasiparticle numbers $N$, and as \textit{topologically inhomogeneous} when they vary with $N$.
This general mechanism will be explicitly illustrated for the AKLT model in Sec.~\ref{sec:AKLT-scar}. 

\begin{table*} 
\caption{
Topological properties of the ground-state MPS, the quasiparticle excitation MPO, and the resulting bimagnon scar subspace for each protecting symmetry (inversion, time reversal, and on-site $\mathbb{Z}_2\times\mathbb{Z}_2$). The scar subspace is homogeneous or inhomogeneous depending on whether the corresponding symmetry induces a projective or linear action that varies with the quasiparticle number $N$.
The boldface entries correspond to the regime studied in this work.
In this regime, the scar subspace is topologically inhomogeneous under inversion, whereas it remains topologically homogeneous under on-site $\mathbb{Z}_2\times\mathbb{Z}_2$ and time-reversal symmetries.
} 
\label{tab:topo_prop}
\begin{center}
\begin{tabular}{| l | c | c | c | c |}
\hline
Ground state MPS & \multicolumn{2}{c|}{trivial} & \multicolumn{2}{c|}{\textbf{nontrivial}} \\ \hline
Excitation MPO & trivial & nontrivial & \textbf{trivial} & \textbf{nontrivial} \\ \hline
Scar subspace & \begin{tabular}{c} topologically \\ homogeneous (trivial) \end{tabular} & \begin{tabular}{c} topologically \\ inhomogeneous \end{tabular} & \begin{tabular}{c} \textbf{topologically} \\ \textbf{homogeneous (nontrivial)} \end{tabular} & \begin{tabular}{c} \textbf{topologically} \\ \textbf{inhomogeneous} \end{tabular} \\ 
\hline
\end{tabular}
\end{center}
\end{table*}

\section{SPT-like Scar Subspace}
\label{sec:SPT-scar}

In this section, we introduce the notion of \added{an SPT-like scar subspace, namely a symmetry-invariant subspace in which topological properties can be consistently defined and analyzed. 
Such subspaces naturally arise in systems where the Hamiltonian is invariant under the protecting symmetries and admits rSGAs. }

We first \added{explain how to} extend the notion of invariance under protecting symmetries from individual states to a subspace, reviewing quasiparticle-based QMBS as a natural framework.
We then show that \added{a symmetry-invariant scar subspace provides a natural setting in which symmetry-associated topological properties are consistently defined.
These symmetry-associated topological properties are inherited from those of the underlying ground-state MPS. }

\subsection{QMBS as quasiparticle excitations}

Quantum many-body scars (QMBS)~\cite{Bernien2017, Shiraishi2017, Turner2018,Turner2018b, Serbyn2021, Moudgalya2022, Chandran2023} were originally introduced as exceptional energy eigenstates that violate the eigenstate thermalization hypothesis (ETH)~\cite{Deutsch1991, Srednicki1994, Rigol2008}. 
In a thermalizing system, these states remain distinguished from thermal states by macroscopic variables. Typically, QMBS form a small invariant subspace of the Hamiltonian. As the scar subspace has vanishing dimension in the thermodynamic limit, generic initial states thermalize, while those overlapping strongly with it avoid thermalization~\cite{Turner2018}. 

The emergence of such invariant subspaces often stems from extra algebraic structures that exist only within them. A prominent example is the restricted spectrum generating algebra (rSGA)~\cite{Moudgalya:2020PhRvB.102h5140M}:
\begin{align} \label{eq:rSGA}
    \bigl([H,\,Q^{\dag}] - \mathcal{E} Q^{\dag}\bigr) \big|_W = 0,   
\end{align}
where $W$ denotes the subspace, and $Q^{\dag}$ is typically a sum of locally supported operators~\cite{Buca2023UnifiedDynamics}.
In the cases of interest, the invariant subspace $W$ is spanned by \add{the quasiparticle states $|\Psi_N \rangle$ introduced in \eqref{eq:q-state}, with $|\Psi_0\rangle$ serving as the reference eigenstate satisfying }$H|\Psi_0\rangle = E_0 |\Psi_0\rangle$.  
This construction captures the most commonly studied form of QMBS towers in the literature~\cite{Moudgalya:2018PhRvB..98w5156M,ODea:2020PhRvR...2d3305O,
Moudgalya2022,Chandran2023}.  

Since $Q^{\dagger}$ is a sum of local operators, the resulting states can be viewed as quasiparticle
excitations, with $|\Psi_N\rangle$ interpreted as an $N$-particle non-interacting quasiparticle state built on top of the reference eigenstate~\cite{Moudgalya:2018PhRvB..98w5156M,ODea:2020PhRvR...2d3305O,Moudgalya2022,Chandran2023,Matsui2024}.
Consequently, the resulting QMBS admit an MPS description with bond dimension not exceeding $D\,D_M^{\,N}$~\cite{Vidal2003,PerezGarcia2007,Schuch2008}, where $D_M$ denotes the bond dimension of the quasiparticle creation MPO. Provided that each QMBS is invariant under the protecting symmetries, this places the rSGA-generated quasiparticle towers precisely within the MPS-based framework developed in the previous section for defining the topological properties of quasiparticle states.

\subsection{Symmetry-invariant subspaces and topological properties}
\label{subsec:inv_subspace}

Individual QMBS are not necessarily invariant under the protecting symmetries, which makes it unclear how their symmetry-associated topological properties should be defined on a state-by-state basis. 
However, as we will show below, rSGA-generated towers span a subspace that \emph{is} invariant under the relevant symmetries.  
Once such a symmetry-invariant scar subspace is identified, its symmetry-associated topological properties can be defined in a consistent manner using the MPS-based framework developed in the previous section. This motivates \added{referring to such subspaces as \textit{SPT-like scar subspaces}, characterized by symmetry-associated topological properties. }

The notion of symmetry—whether on-site, inversion, or time-reversal—extends naturally to subspaces. A subspace $W$ is called $\mathcal{S}$-invariant if it is closed under the action of any protecting symmetry $\mathcal{S}$, i.e.,
\begin{align}
    \mathcal{S}\, W \subseteq W, \quad
    \mathcal{S} \in \{g \in G,\, \mathcal{P},\, \mathcal{T}\},
\end{align}
where $g$ denotes an element of the on-site symmetry group $G$, and each $\mathcal{S}$ acts on the Hilbert space via its (anti-)unitary representation. Individual states in $W$ need not be invariant under $\mathcal{S}$ for the subspace itself to possess this property. 

Consider a Hamiltonian and a reference state that are invariant under each protecting symmetry $\mathcal{S} \in \{g \in G,\, \mathcal{P},\, \mathcal{T}\}$. If a ladder operator $Q^\dagger$ generates a tower of QMBS, its symmetry-transformed operator 
\begin{align}
    Q_{\mathcal{S}}^\dagger := 
    \mathcal{S}\, Q^\dagger\, \mathcal{S}^{-1},
\end{align}
also generates a tower, since it satisfies the rSGA
\begin{align}
    \Big([H,\, Q_{\mathcal{S}}^{\dagger}] - \mathcal{E} Q_{\mathcal{S}}^{\dagger}\Big)
    \Big|_{W_{\mathcal{S}}} = 0,
\end{align}
within the subspace $W_{\mathcal{S}} = \mathrm{span}\{(Q_{\mathcal{S}}^\dagger)^N|\Psi_0\rangle\}$. 
Collecting all such towers related by symmetry operations, we define the symmetry-invariant subspace
\begin{align}
    W_{\rm sym} := \mathrm{span}\bigcup_{\mathcal{S} \in \{g \in G,\, \mathcal{P},\, \mathcal{T}\}} W_{\mathcal{S}},
\end{align}
which is closed under the action of the protecting symmetries and hence satisfies $\mathcal{S} W_{\rm sym} \subseteq W_{\rm sym}$ for all $\mathcal{S} \in \{g \in G,\, \mathcal{P},\, \mathcal{T}\}$.

For concreteness, we focus on the abelian group $G = \mathbb{Z}_2 \times \mathbb{Z}_2$ generated by two distinct $\mathbb{Z}_2$ elements $g_x$ and $g_z$.  
If the quasiparticle creation operators $Q^\pm$ are chosen to satisfy
\begin{align} \label{eq:Ginv-creation}
    &U_{g_z} Q^\pm U_{g_z}^\dag = Q^\pm, \\
    &U_{g_x} Q^\pm U_{g_x}^\dag = Q^\mp, \nonumber
\end{align}
where $U_{g_x}$ and $U_{g_z}$ denote the global physical unitary representations of $g_x$ and $g_z$, respectively, then the resulting $\mathbb{Z}_2 \times \mathbb{Z}_2$-invariant scar subspace $W_{\mathbb{Z}_2 \times \mathbb{Z}_2}$ is formed by two QMBS towers,
\begin{align}
    |\Psi_N^\pm \rangle = (Q^\pm)^N |\Psi_0 \rangle,
\end{align}
corresponding to two species of quasiparticle excitations.

\added{When the subspace is symmetry-invariant, symmetry-associated topological properties can be consistently defined across the scar subspace, even if individual quasiparticle states are not invariant under the symmetry. These properties are encoded in the bond-space representations of the corresponding quasiparticle MPS. }

The MPO tensors associated with the quasiparticle creation operators $Q^\pm$ transform under the local physical $\mathbb{Z}_2 \times \mathbb{Z}_2$ actions $u_{g_z}$ and $u_{g_x}$ as
\begin{align}
    &u_{g_z} M_\pm u_{g_z}^\dag = 
        e^{i\theta_{g_z}^\pm} V_{g_z}^\pm M_\pm (V_{g_z}^\pm)^\dag, \\
    &u_{g_x} M_\pm u_{g_x}^\dag = 
        e^{i\theta_{g_x}^\pm} V_{g_x}^\pm M_\mp (V_{g_x}^\pm)^\dag.   
\end{align}
For the \added{symmetry-associated} topological \added{properties} to be well defined, the two bond-space representations $\{V_g^+\}$ and $\{V_g^-\}$ must belong to the same projective class in $H^2(G,U(1))$.  
This requirement is satisfied when the two representations are related by conjugation with a group element (an inner automorphism) together with a change of basis, which is precisely the case for any symmetry-invariant scar subspace. See Appendix~\ref{ap:Lemma} for a short proof. 

A sufficient condition \added{for this consistency} is that the two representations are identical at the matrix level,
\begin{align}
   V_g^+ = V_g^- = V'_g ,
\end{align}
which indeed holds for the bimagnon scar subspace of the AKLT model, as 
shown later in this paper.
With this topological consistency condition, all QMBS in the symmetry-invariant scar subspace share topological characteristics that depend only on the quasiparticle number $N$, not on their species.

Depending on whether the MPO tensors $M_\pm$ transform linearly or projectively under the protecting symmetries, the scar subspace may be topologically homogeneous or inhomogeneous (Table~\ref{tab:topo_prop}).  
When $M_\pm$ transform linearly, the resulting bond representation $\mathbb{V}'_{g} \otimes_a V_{g}$ ($\mathbb{V}'_{g} = (V'_g)^{\otimes_b N}$) carries the same topological character as that of the reference state, yielding a topologically homogeneous scar 
subspace.  
In contrast, a projective transformation of $M_\pm$ leads to a topologically inhomogeneous scar subspace, in which the bond representation $\mathbb{V}'_{g} \otimes_a V_{g}$ exhibits a topological character that depends on the quasiparticle number.

In this way, a symmetry-invariant scar subspace provides a platform \added{in which symmetry-associated topological properties can be consistently defined}. We therefore refer to such a subspace as an \emph{SPT\added{-like} scar subspace}. 
In the next section, we demonstrate these \removed{concepts }\added{structures} concretely in the AKLT model.

\section{Topological properties of the AKLT scar subspace}
\label{sec:AKLT-scar}

In this section, we investigate the topological properties of the scar subspace in the Affleck–Kennedy–Lieb–Tasaki (AKLT) model~\cite{Affleck1987, Affleck1988}. 
The AKLT chain serves as an ideal testing ground for our framework, as it possesses an exactly solvable SPT matrix-product ground state together with a tower of quasiparticle-like QMBS built on top of it~\cite{Moudgalya:2018PhRvB..98w5156M,Chandran2023,Matsui2024}. \added{We also demonstrate how these topological properties give rise to observable signatures in the scar subspace. }

Building on the SPT\added{-like} scar-subspace framework developed in the previous section, we demonstrate that the scar subspace of the AKLT model may either retain or modify the \removed{symmetry-protected topological (SPT) }\added{symmetry-associated topological} \added{properties} of the ground state, depending on the protecting symmetry~\cite{Pollmann2010,Pollmann2012,Schuch2011,ChenGuWen2011}. 
This is established through complementary analyses of the symmetry representations in the matrix-product representation, \added{expectation values of symmetry-twist operators}, and \added{analytically established }long-range string order \added{in the dilute regime}, supplemented by numerical \added{evidence showing behavior distinct from typical thermal states}~\cite{denNijs1989,KennedyTasaki1992,Zaletel:2014JSMTE..10..007Z}. 

By comparing these quantities between the ground state and the scar subspace, we establish that the latter may either faithfully inherit or modify \added{the symmetry-associated topological properties} of the former, depending on the protecting symmetry.

\subsection{Quantum many-body scars in the AKLT model}

The AKLT model~\cite{Affleck1987,Affleck1988} is a paradigmatic non-integrable spin-$1$ spin chain whose Hamiltonian is defined as
\begin{align}  
    &H = \sum_{j=1}^L h_{j,j+1}, \nonumber \\
    &h_{j,j+1} = \vec{S}_j \cdot \vec{S}_{j+1} + \frac{1}{3} (\vec{S}_j \cdot \vec{S}_{j+1})^2,  \label{eq:AKLT}
\end{align}
\added{where $h_{j,j+1}$ is proportional to the projector onto the total spin-$2$ subspace, up to normalization and a constant shift. }
Here $\vec{S}_j = {}^{t}(S_j^{x}, S_j^{y}, S_j^{z})$ denotes the operator-valued vector whose components are the spin-1 operators acting on the $j$th site.
The model possesses a unique ground state that can be exactly represented as an MPS~\cite{Affleck1987, Affleck1988}:
\begin{align} \label{eq:AKLT_gs}
    &|\Psi_0 \rangle = {\rm tr}_{\rm a}(\vec{A} \otimes_p \vec{A} \otimes_p \cdots \otimes_p \vec{A}), \nonumber \\
    &\vec{A} = \begin{pmatrix}
        \sqrt{\frac{2}{3}}\,\sigma_a^+, &
        -\sqrt{\frac{1}{3}}\,\sigma_a^z, &
        -\sqrt{\frac{2}{3}}\,\sigma_a^-
    \end{pmatrix}. 
\end{align}
The Hamiltonian is symmetric under the on-site $\mathbb{Z}_2 \times \mathbb{Z}_2$ spin-rotation symmetry, spatial inversion $\mathcal{P}$, and time-reversal symmetry $\mathcal{T}$, and its ground state inherits the same set of symmetries~\cite{Pollmann2010, Pollmann2012}.

Beyond the ground state, the AKLT Hamiltonian hosts a set of exact excited eigenstates, which admit a quasiparticle interpretation~\cite{Arovas1989, Moudgalya2022, Mark2020, Lin2020, Matsui2024}, including the well-known Arovas states~\cite{Arovas1989}. 
In particular, two independent towers of bimagnon scar states arise from the following rSGAs~\cite{Moudgalya:2020PhRvB.102h5140M}:
\begin{align} 
    &\bigl([H,\,Q^{\pm}] - 2 Q^{\pm}\bigr) \big|_{W_{\pm}} = 0, \nonumber \\
    &Q^{\pm} = \sum_{j=1}^L (-1)^j (S_j^{\pm})^2. \label{eq:rSGA_AKLT}
\end{align}
The corresponding subspaces $W_{\pm}$ are spanned by QMBS built on 
top of the ground state $|\Psi_0\rangle$, 
\begin{align}
|\Psi^{\pm}_N \rangle = (Q^{\pm})^N |\Psi_0 \rangle.
\end{align}

The two towers generated by $Q^+$ and $Q^-$ are separately invariant under $R^z$, forming a $\mathbb{Z}_2$ subgroup of $\mathbb{Z}_2 \times \mathbb{Z}_2$, while the generator $R^x$ exchanges the two towers. Consequently, the combined subspace $W = \mathrm{span}(W_+ \cup W_-)$ is invariant under the on-site $\mathbb{Z}_2 \times \mathbb{Z}_2$ symmetry generated by $R^x$ and $R^z$.
In addition, the translationally invariant structure of the bimagnon creation operator ensures that the subspace $W$ is invariant under inversion.  Moreover, $W$ is also invariant under time reversal $\mathcal{T} = R^y \circ \mathcal{K}$, where $\mathcal{K}$ denotes complex conjugation, since it consists of quasiparticle states with momentum $\pi$ built on the $R^y$-invariant ground state.

\subsection{Bond-space symmetry representations of the scar subspace}\label{subsec:bond_rep}

As shown above, \added{the scar subspace, including the ground state,} is invariant inversion, time reversal, and the on-site $\mathbb{Z}_2\times\mathbb{Z}_2$ symmetries. Since the ground state is represented by an injective MPS, each symmetry induces a corresponding transformation on the bond space~\cite{Sanz:2009PhRvA..79d2308S, Schuch2011, Pollmann2010, Pollmann2012} as 
\begin{align}
    &\mathcal{S}:~ \vec{A} \mapsto 
    e^{i\theta_{\mathcal{S}}}\,
    V_{\mathcal{S}}\, \vec{A}\, V_{\mathcal{S}}^{\dag},
    \\
    &\mathcal{S}\in\{\mathcal{P},\,\mathcal{T},\,g_x,\,g_z\}. \nonumber
\end{align}
The matrices $V_{\mathcal{S}}$ furnish projective representations on the bond space~\cite{Pollmann2010, Pollmann2012}:
\begin{align} \label{eq:rep_gs}
    &V_{\mathcal{P}} = i\sigma^y, \\
    &V_{\mathcal{T}} = i\sigma^y, \nonumber\\
    &V_{g_\alpha} = \sigma^{\alpha} \quad (\alpha = x,z). \nonumber
\end{align}
Therefore, the AKLT ground state \added{exhibits nontrivial topological properties associated with} $\mathcal{P}$, $\mathcal{T}$, or $\mathbb{Z}_2 \times \mathbb{Z}_2$~\cite{Pollmann2010, Pollmann2012}\added{, consistent with its well-known SPT character}.

In the following, we analyze how each of these symmetries acts on the bimagnon scar subspace and show that its topological properties are either faithfully inherited from the ground state or modified in a symmetry-dependent manner.

\subsubsection{Inversion symmetry}

As discussed above, the two species of bimagnon QMBS are individually invariant under inversion.
More explicitly, inversion acts on the $N$-bimagnon QMBS only through an overall phase,
\begin{align}
    \mathcal{P}:~|\Psi_N^{\pm}\rangle \mapsto (-1)^{N(L-1)}|\Psi_N^{\pm}\rangle,
\end{align}
which is physically irrelevant, so each QMBS is effectively inversion-invariant.

The local inversion operator transposes both MPS and MPO tensor elements 
in the bond space, with the explicit action on the bimagnon creation MPO
\begin{align}
    \mathcal{P}:~
    M^{\pm} 
    \longmapsto {^{t_b}}(M^{\pm})
    = - V'_{\mathcal{P}}\, M^{\pm}\, V_{\mathcal{P}}'^{\dag}.
\end{align}
In the bimagnon QMBS of the AKLT model, this transformation is realized by the unitary bond operator $V'_{\mathcal{P}} = i\sigma^y$, as noted in \cite{Moudgalya:2018PhRvB..98w5156M}.  
Together with the ground-state MPS bond action $V_{\mathcal{P}} = i\sigma^y$ in \eqref{eq:rep_gs}, inversion of an $N$-bimagnon QMBS is represented on the bond space by $V_{\mathcal{P}} \otimes \mathbb{V}’{\mathcal{P}}$, where $\mathbb{V}’{\mathcal{P}} = (V_{\mathcal{P}}’)^{\otimes_b N}$. This yields a linear representation of inversion for odd $N$ and a projective one for even $N$. 

Hence, the scar subspace exhibits trivial and nontrivial topological properties in an alternating fashion as the number of bimagnons $N$ changes between odd and even.

\subsubsection{Time-reversal symmetry}

Unlike inversion, the individual bimagnon QMBS are not invariant under time reversal. Instead, the time-reversal operator connects the bimagnon QMBS of different species as 
\begin{align} 
    \mathcal{T}:~ |\Psi_N^{\pm}\rangle 
    \mapsto |\Psi_N^{\mp}\rangle, 
\end{align} 
and thus the scar subspace as a whole is invariant under time reversal. 
Since the time-reversal operator locally acts as $\mathcal{T} = R^{y} \circ \mathcal{K}$ in the spin-$1$ AKLT model, the MPO elements transform as
\begin{align}
    \mathcal{T}:~ M^{\pm}
    \mapsto V'_{\mathcal{T}}\, M^{\mp}\, V_{\mathcal{T}}'^{\dag},
\end{align}
where, for the bimagnon QMBS, the unitary bond transformation $V'_{\mathcal{T}}$ is realized as the identity operator.
 
As a result, the bond transformation of the $N$-bimagnon QMBS under time reversal, $V_{\mathcal{T}} \otimes \mathbb{V}'_{\mathcal{T}}$, forms a projective representation for any number of quasiparticles $N$, indicating that the quasiparticle creation operator does not alter the 
topological character of the ground state. Therefore, the topological \added{properties} of the ground state \added{associated with time-reversal symmetry} is faithfully inherited within the bimagnon scar subspace.

\subsubsection{$\mathbb{Z}_2 \times \mathbb{Z}_2$ symmetry}

The $\mathbb{Z}_2 \times \mathbb{Z}_2$ symmetry, generated by $g_x$ and $g_z$, acts on spin-$1$ sites as $\pi$ rotations about the $x$ and $z$ axes, denoted by $R^x$ and $R^z$, respectively.
The bimagnon QMBS are invariant under $R^z$, while $R^x$ exchanges the quasiparticle species:
\begin{align}
    &g_z:\quad |\Psi_N^{\pm}\rangle \mapsto |\Psi_N^{\pm}\rangle, \\
    &g_x:\quad |\Psi_N^{\pm}\rangle \mapsto |\Psi_N^{\mp}\rangle. 
\end{align}
Hence, the bimagnon scar subspace $W$ is invariant under both $R^{z}$ and $R^{x}$, and therefore invariant under the full on-site $\mathbb{Z}_{2} \times \mathbb{Z}_{2}$ symmetry.

Using the standard spin-1 representations of $R^{x}$ and $R^{z}$, the bimagnon creation MPO tensors transform as
\begin{align}
    &R^{z} M^{\pm} (R^{z})^{\dag} 
        = V'_{g_{z}}\, M^{\pm}\, V_{g_{z}}'^{\dag}, \\
    &R^{x} M^{\pm} (R^{x})^{\dag} 
        = V'_{g_{x}}\, M^{\mp}\, V_{g_{x}}'^{\dag}, \nonumber
\end{align}
where both unitary bond transformations $V'_{g_{z}}$ and $V'_{g_{x}}$ are realized as identities.
Thus, for the $N$-bimagnon QMBS, the $\mathbb{Z}_{2} \times \mathbb{Z}_{2}$ generators act on the bond space as $V_{g_{\alpha}} \otimes \mathbb{V}'_{g_{\alpha}}$ ($\alpha = z, x$), forming projective representations for any number of quasiparticles $N$.
Consequently, the scar subspace composed of the bimagnon QMBS preserves the same topological properties as the ground state \added{with respect to the on-site $\mathbb{Z}_2 \times \mathbb{Z}_2$ symmetry}.

Among the symmetries of the bimagnon scar subspace, the on-site $\mathbb{Z}_{2} \times \mathbb{Z}_{2}$ symmetry gives rise to \added{observable signatures} of nontrivial topology.  
In the following subsections, we illustrate these signatures through the analysis of \added{expectation values of symmetry-twist operators} and long-range string order.

\subsection{Expectation values of symmetry-twist operators}

A hallmark of nontrivial \added{symmetry-associated topological properties} is \added{a projective symmetry representation in the bond space of MPS, which can be detected through expectation values of symmetry-twist operators}~\cite{Chen2012Science,Zaletel:2014JSMTE..10..007Z}. 
Within the AKLT scar subspace, \added{these signature} can be directly probed through phase factors generated by twisting the \added{associated} symmetry at the boundary.

Before turning to the AKLT case, we briefly recall \added{the standard construction of symmetry-twist operators for the ground state, often referred to as a topological response}.
Suppose that $|\Psi_0\rangle$ is an MPS ground state \added{associated with} an on-site symmetry group $G$.
With $V_{g_k}$ representing the (possibly projective) action of $g_k \in G$ on the bond space, \added{a phase factor associated with} another group element $g_j$ is defined by~\cite{Zaletel:2014JSMTE..10..007Z}
\begin{align} \label{eq:topo_res_gs_fixed}
    e^{i\phi_{g_k,g_j}} =
    \frac{\langle \Psi_0(V_{g_k})|\, u_{g_j}^{\otimes L}\, |\Psi_0(V_{g_k}) \rangle}
    {\langle \Psi_0|\, u_{g_j}^{\otimes L}\, |\Psi_0 \rangle},
\end{align}
where $|\Psi_0(V_{g_k})\rangle$ denotes the MPS with a boundary twist,
\begin{align}
    |\Psi_0(V_{g_k})\rangle
    = \sum_{m_1,\dots,m_L} {\rm tr}(V_{g_k} A_{m_1}\cdots A_{m_L})\, |m_1\dots m_L\rangle.
\end{align}
For an injective MPS, this ratio is determined solely by the projective phases of the bond representation. 
From the projective multiplication laws \eqref{eq:proj} and \eqref{eq:commuting_phase}, 
it follows that 
\begin{align} \label{eq:commutator_phase}
    e^{i\phi_{g_k,g_j}} = \frac{\omega(g_k,g_j)}{\omega(g_j,g_k)} 
    = e^{i\phi(g_k,g_j)}.
\end{align}
\added{The associated phase $\phi(g_k,g_j)$ is thus determined by the commutation relation of $V_{g_k}$ and $V_{g_j}$ at the boundary. }
In particular, for $G=\mathbb{Z}_2\times\mathbb{Z}_2$, \added{this phase takes the value} $\pm 1$ depending on the corresponding $2$-cocycle. 

In the AKLT model, a standard choice is $g_k = g_x$ and $g_j = g_z$, whose projective commutation relation yields $\phi_{g_x,g_z} = \phi(g_x,g_z) = \pi$~\cite{Pollmann2010, Pollmann2012}.  
We now \added{extend this symmetry-twist construction} to the bimagnon scar states. 
Using their MPS representations, we define
\begin{align} \label{eq:topo_res_bi}
    e^{i\widetilde{\phi}_{g_x,g_z}} =
    \frac{\langle \Psi^{\pm}_N(V_{g_x})| \prod_{k=1}^{L} R_k^z |\Psi^{\pm}_N(V_{g_x}) \rangle}
         {\langle \Psi^{\pm}_N| \prod_{k=1}^{L} R_k^z |\Psi^{\pm}_N \rangle},
\end{align}
where $|\Psi^{\pm}_N(V_{g_x})\rangle$ denotes the bimagnon state with a boundary twist implemented by $V_{g_x}$,
\begin{align}
    &|\Psi^{\pm}_N(V_{g_x}) \rangle
    = (Q^{\pm}(V'_{g_x}))^N |\Psi_0(V_{g_x}) \rangle, \\
    &Q^{\pm}(V'_{g_x}) = {_b}\langle \uparrow\!| V'_{g_x} \cdot (M^{\pm})^{\otimes_{\rm p} L} |\!\downarrow \rangle_b. \nonumber
\end{align}
Here, the boundary operators $V_{g_x}$ and $V'_{g_x}$ \added{acting on the bond space} are chosen consistently with the bond representations of $g_x$, namely $V_{g_x} \otimes \mathbb{V}'_{g_x} = \sigma^x \otimes \bm{1}$. 
The bulk transformation by $R^z$ acts as $V_{g_z} \otimes \mathbb{V}'_{g_z} = \sigma^z \otimes \bm{1}$ on the bond space.  
Consequently, the phase factor $e^{i\widetilde{\phi}_{g_x,g_z}}$ again takes the nontrivial value $-1$, \added{encoded in} the projective representations of the bond transformations $V_{g_\alpha} \otimes \mathbb{V}'_{g_\alpha}$ ($\alpha=x,z$) associated with the bimagnon MPS.

Therefore, the bimagnon scar subspace of the AKLT model exhibits the same nontrivial \added{phase signatures} as the ground state, demonstrating that the \added{symmetry-associated topological properties} of the ground state \added{are} faithfully \added{inherited by} the bimagnon scar subspace. 
This observation illustrates how \added{symmetry-associated} topological properties can persist throughout a \added{scar} subspace \added{generated by rSGAs and therefore invariant under the underlying symmetry}.

\subsection{Long-range string order}

Another key manifestation of \added{symmetry-associated topological properties} is the presence of finite long-range string order, which encodes hidden structure invisible to local observables~\cite{denNijs1989, KennedyTasaki1992, Turner2011}. 
For the AKLT model, the nonvanishing string order parameter in the ground state serves as a hallmark of its nontrivial \added{symmetry-associated topological properties}~\cite{Pollmann2010,Pollmann2012}. 
We now demonstrate that string order \added{asymptotically approaches a fixed value} across the entire bimagnon scar subspace \added{in the dilute regime}.

For an energy eigenstate $|\Psi\rangle$ of a Hamiltonian that is symmetric under an on-site symmetry group $G$, string order is defined through the nonlocal correlator associated with a local operator $\chi$ and a group element $g_k \in G$ as
\begin{align} 
    O_{\rm string}(\ell_1,\ell_2;\chi,g_k)
    = \frac{\langle \Psi|
       \chi_{\ell_1} \cdot 
       \prod_{\ell_1<\ell<\ell_2} u_{g_k,\ell} \cdot 
       \chi_{\ell_2}
       |\Psi \rangle}
       {\langle \Psi|\Psi\rangle}. 
    \label{eq:string_order_def} 
\end{align}
Taking the thermodynamic limit first and then the large-distance limit, this correlator defines the so-called string order parameter, which characterizes hidden long-range order.
Each local operator $\chi$ is assumed to carry a definite charge $\sigma_j$ with respect to $g_j$, defined through
\begin{align}
    u_{g_j}\chi u_{g_j}^\dagger = e^{i\sigma_j}\chi.
\end{align}

In general, a nonzero asymptotic value of \eqref{eq:string_order_def} signals nontrivial \added{symmetry-associated topological properties} in eigenstates $|\Psi\rangle$ \added{that respect the underlying on-site symmetry}. 
When $|\Psi\rangle$ is a translationally invariant MPS with on-site symmetry $G$, the string order can remain finite only when the phase $\sigma_j$ associated with the local operator $\chi$ matches the projective phase $\phi(g_k,g_j)$ \eqref{eq:commuting_phase}.

For the AKLT model, the standard choice $\chi = S^z$, $u_{g_k} = R^z$, and $u_{g_j} = R^x$ yields $\sigma_z = \phi_{g_x,g_z} = \pi$, leading to a finite long-range string order~\cite{denNijs1989, KennedyTasaki1992, Pollmann2010, Pollmann2012}. 
Below, we first review its evaluation for the ground state and then extend the analysis to the bimagnon QMBS.

\subsubsection{String order in the ground state}

Expectation values of physical operators on an MPS are conveniently expressed using the transfer-matrix formalism~\cite{Fannes1992}. 
For an operator $X$ acting on the physical space, we define the generalized transfer matrix (which reduces to the standard transfer matrix for $X=\bm{1}$) as
\begin{align}
    T_X := {^{t_p}\!}\vec{A}_{a'}^{*} X \vec{A}_a .
\end{align}
Accordingly, the norm of the AKLT ground state can be written as
\begin{align} \label{eq:norm_gs}
    \langle \Psi_0|\Psi_0 \rangle 
    = \mathrm{tr}_{a',a}\!\left(T_{\bm{1}}^{L}\right).
\end{align}
The numerator of the string correlator \eqref{eq:string_order_def} can also be expressed using the generalized transfer matrices as
\begin{align} \label{eq:SO_gs}
    &\langle \Psi_0| S^z_{\ell_1} \cdot \prod_{\ell = \ell_1+1}^{\ell_2-1} R_{\ell}^z \cdot S^z_{\ell_2} |\Psi_0 \rangle \nonumber\\
    &= {\rm tr}_{a',a}\,\bigl( 
       T_{\bm{1}}^{\ell_1-1} \, T_{S^z} \, T_{R^z}^{\ell_2-\ell_1-1} \, T_{S^z} \, T_{\bm{1}}^{L-\ell_2}\bigr).
\end{align}

\added{
The AKLT ground-state MPS is injective, and the transfer matrix $T_{\bm{1}}$ therefore has a unique eigenvalue of largest magnitude $\lambda_{\rm max} = 1$ with eigenvector $|\lambda_{\rm max} \rangle$.
The projective bond-space representations of the $\mathbb{Z}_2 \times \mathbb{Z}_2$ symmetry imply the relation $T_{R^{z}} = \sigma_{a}^{z} T_{\bm{1}} \sigma_{a}^{z}$.
In the thermodynamic limit, both numerator and denominator are dominated by the contribution of this leading eigenvector.
}

\added{
With these observations, the long-range correlator on the AKLT ground state,
$O^{(0)}_{\rm string}(\ell_1,\ell_2;S^z,g_z)$, approaches a finite value
at large system size and separation~\cite{denNijs1989}:
\begin{align} \label{eq:string_gs}
    &O^{(0)}_{\rm string}(\ell_1,\ell_2;S^z,g_z) \nonumber \\
    &= \lambda_{\rm max}^{-2}\,
      \langle \lambda_{\rm max}|
      T_{S^z}\sigma_a^z |\lambda_{\rm max} \rangle 
      \langle \lambda_{\rm max}| \sigma_a^z T_{S^z} |\lambda_{\rm max} \rangle + O\Big( \Big( \frac{1}{3} \Big)^L \Big) \nonumber \\
    &= -\frac{4}{9} + O\Big( \Big( \frac{1}{3} \Big)^L \Big),
\end{align}
which defines the string order parameter $O^{(0)}_{\rm string}$ by the limits  
\begin{align}
    O^{(0)}_{\rm string}
    = \lim_{|\ell_2-\ell_1|\to\infty} \lim_{L\to\infty} \, O^{(0)}_{\rm string}(\ell_1,\ell_2;S^z,g_z).
\end{align} 
}

In contrast, the ordinary two-point correlation \added{$\mathcal{O}_{zz}(\ell_1,\ell_2) = \langle \Psi| S^z_{\ell_1} S^z_{\ell_2} |\Psi \rangle /  \langle \Psi|\Psi\rangle$} in the AKLT ground state decays exponentially with distances as \added{$\mathcal{O}_{zz}(\ell_1,\ell_2) \sim (-1/3)^{|\ell_2-\ell_1|}$}~\cite{Affleck1988,Fannes1992}.

\subsubsection{String order in the bimagnon scar subspace}

We now turn to the bimagnon scar subspace and show that\added{, in the dilute regime,} the long-range string order persists throughout the bimagnon scar tower, attaining the same value as in the ground state.

For simplicity, let us start with the single-bimagnon excitation ($N=1$). In this case, the correlation functions are expressed using the tensored generalized transfer matrices
\begin{align} \label{eq:single_bimagnon}
    \mathbb{T}^{\pm\pm}_X &= {^{t_p}\!}\vec{A}_{a'}^{*}\, {^{t_p}\!}M_{b'}^{\pm *} X M_b^{\pm}\, \vec{A}_a \nonumber \\
    &= \sigma^z_{b'} \sigma^z_b\, {^{t_p}\!}\vec{A}_{a'}^{*} X \vec{A}_a
    + \sigma^+_{b'} \sigma^+_b\, {^{t_p}\!}\vec{B}_{a'}^{*} X \vec{B}_a \nonumber \\
    &+ \sigma^z_{b'} \sigma^+_b\, {^{t_p}\!}\vec{A}_{a'}^{*} X \vec{B}_a
    + \sigma^+_{b'} \sigma^z_b\, {^{t_p}\!}\vec{B}_{a'}^{*} X \vec{A}_a. 
\end{align}
The second equality follows from expanding the MPO tensors $M_b^{\pm}$ in the physical space, introducing $\vec{B} := (S^+)^2 \vec{A}$ at the positions of bimagnon excitations. 
Accordingly, the norm reads
\begin{align} \label{eq:norm_bi}
    \langle \Psi_1^{\pm}|\Psi_1^{\pm} \rangle
    = {\rm tr}_{b',a',a,b}\,
      \Big(K_{b',b} (\mathbb{T}^{\pm\pm}_{\bm{1}})^L\Big),
\end{align}
where $K_{b',b} = \sigma_{b'}^- \sigma_b^-$ encodes the boundary conditions on the bond space of the bimagnon-creation MPO.

\added{
Due to the boundary condition $K_{b',b}=\sigma^-_{b'}\sigma^-_b$, only terms containing exactly one $\sigma^+$ operator on each of the bond spaces $\mathcal{V}_b$ and $\mathcal{V}_{b'}$ survive in the expansion of $(\mathbb{T}^{\pm\pm}_{\bm 1})^L$. 
The resulting contributions are restricted to configurations containing either $L-1$ factors of $({^{t_p}\!}\vec{A}_{a'}^* \vec{A}_a)$ and one factor of $({^{t_p}\!}\vec{B}_{a'}^* \vec{B}_a)$, or $L-2$ factors of $({^{t_p}\!}\vec{A}_{a'}^* \vec{A}_a)$ together with one factor each of $({^{t_p}\!}\vec{A}_{a'}^* \vec{B}_a)$ and $({^{t_p}\!}\vec{B}_{a'}^* \vec{A}_a)$, with arbitrary ordering along the chain.
In both cases, the number of $\vec{B}$ insertions remains finite, so that in the large-$L$ limit the behavior is governed by the dominant eigenvalue $\lambda_{\rm max}$ of $T_{\bm{1}} = {^{t_p}}\vec{A}_{a'}^* \vec{A}_a$ at large $L$.
}

\added{
The former pairs contribute with a multiplicity proportional to $L$, while the latter consists of $O(L^2)$ distinct configurations. However, the $O(L^2)$ contribution vanishes due to the condition
\begin{align} \label{eq:vanishing_conf}
    \langle \lambda_{\rm max}| {^{t_p}}\vec{A}_{a'}^* \vec{B}_a |\lambda_{\rm max} \rangle = 0.
\end{align}
Therefore, the leading contribution is given by the $O(L)$ term:
\begin{align}
    \langle \Psi_1^{\pm}|\Psi_1^{\pm}\rangle
    = L\, \lambda_{\rm max}^{L-1}\, \alpha_{\rm ex}^{(1)} + o(L),
\end{align}
where $\alpha_{\rm ex}^{(1)} := \langle \lambda_{\rm max}| {^{t_p}}\!\vec{B}_{a'}^* \vec{B}_a |\lambda_{\rm max} \rangle$ is an $O(1)$ coefficient. 
}

On the other hand, the numerator of the string correlator \eqref{eq:string_order_def} can be written as 
\begin{align} \label{eq:string_order_bi}
    &\langle \Psi_1^{\pm}| S^z_{\ell_1}
      \prod_{\ell = \ell_1+1}^{\ell_2-1} R_{\ell}^z
      S^z_{\ell_2} |\Psi_1^{\pm}\rangle \nonumber \\
    &= {\rm tr}_{b',a',a,b}\, \Big(K_{b',b}
       (\mathbb{T}^{\pm\pm}_{\bm 1})^{\ell_1-1}
       \mathbb{T}^{\pm\pm}_{S^z}
       (\mathbb{T}^{\pm\pm}_{R^z})^{\ell_2 - \ell_1 - 1}
       \nonumber \\
    &\hspace{45mm} \cdot \mathbb{T}^{\pm\pm}_{S^z} 
    (\mathbb{T}^{\pm\pm}_{\bm 1})^{L - \ell_2} \Big).    
\end{align}
This expression can be simplified using the local relation $\mathbb{T}^{\pm\pm}_{R^z} = \sigma_a^z \mathbb{T}^{\pm\pm}_{\bm{1}} \sigma_a^z$, which arises from the projective action of the on-site $\mathbb{Z}_2 \times \mathbb{Z}_2$ symmetry on the bond space.

\added{
The dominant contribution in the thermodynamic limit arises, as in the norm, from configurations that survive the boundary condition $K_{b',b}$ and are governed by the leading eigenvalue of $T^{\pm\pm}_{\bm{1}}$, resulting in the same scaling with system size, 
\begin{align}
    &\langle \Psi_1^{\pm}| S^z_{\ell_1}
      \prod_{\ell = \ell_1+1}^{\ell_2-1} R_{\ell}^z
      S^z_{\ell_2} |\Psi_1^{\pm}\rangle \nonumber \\
    &= {\rm tr}_{b',a',a,b}\, \Big( K_{b',b}
       (\mathbb{T}^{\pm\pm}_{\bm 1})^{\ell_1-1}
       \mathbb{T}^{\pm\pm}_{S^z} \sigma^z_a
       (\mathbb{T}^{\pm\pm}_{\bm 1})^{\ell_2 - \ell_1 - 1}
       \nonumber \\
    &\hspace{45mm}\cdot \sigma^z_a \mathbb{T}^{\pm\pm}_{S^z} (\mathbb{T}^{\pm\pm}_{\bm 1})^{L - \ell_2} \Big)
       \nonumber \\
    &= (L-2)\, \lambda_{\rm max}^{L-3}\, |\alpha_{\rm ex}^{(1)}|^2 
    \langle \lambda_{\rm max}| T_{S^z} \sigma^z_a |\lambda_{\rm max} \rangle \nonumber \\
    &\hspace{24mm}\cdot \langle \lambda_{\rm max}| \sigma^z_a T_{S^z} |\lambda_{\rm max} \rangle + o(L), 
\end{align}
where we have used the vanishing condition \eqref{eq:vanishing_conf}. 
With all these, the system-size dependence and the $O(1)$ factor $|\alpha_{\rm ex}^{(1)}|^2$ cancel between the numerator and denominator.
}

\added{
Thus, we find that the single-bimagnon QMBS yields the same asymptotic value of the string correlator as the ground state at large system size and separation: 
\begin{align}
    &O^{(1,\pm)}_{\rm string}(\ell_1,\ell_2;S^z,g_z) \nonumber \\
    &= \lambda_{\rm max}^{-2} \langle \lambda_{\rm max}| T_{S^z} \sigma^z_a |\lambda_{\rm max} \rangle \langle \lambda_{\rm max}| \sigma^z_a T_{S^z} |\lambda_{\rm max} \rangle + O(L^{-1}) \nonumber \\
    &= -\frac{4}{9} + O(L^{-1}), 
\end{align}
which defines the string order parameter in the single-bimagnon excitation state $O^{(1,\pm)}_{\rm string}$ as
\begin{align}
O^{(1,\pm)}_{\rm string}
= \lim_{|\ell_2-\ell_1|\to\infty} \lim_{L\to\infty} \, O^{(1,\pm)}_{\rm string}(\ell_1,\ell_2;S^z,g_z).
\end{align} 
}

For the general $N$-bimagnon QMBS, $|\Psi^{\pm}_N\rangle = (Q^\pm)^N |\Psi_0\rangle$, the string correlator \eqref{eq:string_order_def} can be analyzed in an analogous manner to the ground state and single-bimagnon cases, using the $(N+1)$-fold tensored generalized transfer matrices
\added{
\begin{align}
    \mathbb{T}^{\pm\pm}_X 
    &= {^{t_p}\!}\vec{A}_{a'}^*\, {^{t_p}}\mathbb{M}^{\pm *}_{b'} X \mathbb{M}^{\pm}_b\, \vec{A}_a \nonumber \\
    &= \prod_{j,j'=1}^N \sigma^z_{b'_{j'}}\sigma^z_{b_j}\, {^{t_p}\!}\vec{A}_{a'}^{*} X \vec{A}_a \nonumber \\
    &+ \sum_{k'=1}^N \sigma^+_{b'_{k'}} \prod_{j=1,\, j' \neq k'}^N \sigma^z_{b'_{j'}} \sigma^z_{b_j}\, {^{t_p}\!}\vec{B}_{a'}^{*} X \vec{A}_a \nonumber \\
    &+ \sum_{k=1}^N \sigma^+_{b_{k}} \prod_{j'=1,\, j \neq k}^N \sigma^z_{b'_{j'}} \sigma^z_{b_j}\, {^{t_p}\!}\vec{A}_{a'}^{*} X \vec{B}_a \nonumber \\
    &+ \sum_{k,k'=1}^N \sigma^+_{b_{k}} \sigma^+_{b'_{k'}} \prod_{j \neq k,\,j' \neq k'}^N \sigma^z_{b'_{j'}} \sigma^z_{b_j}\, {^{t_p}\!}\vec{B}_{a'}^{*} X \vec{B}_a. 
\end{align}
}

\added{Due to this transfer-matrix structure, both the denominator and the numerator of the string order correlator are dominated by the eigenvalue of the largest magnitude $\lambda_{\rm max}$ of $\mathbb{T}^{\pm\pm}_{\bm 1}$ in the thermodynamic and long-distance limit. For the numerator, this follows from the local relation $\mathbb{T}^{\pm\pm}_{R^z} = \sigma_a^z \mathbb{T}^{\pm\pm}_{\bm{1}} \sigma_a^z$. Thus, the leading contribution in the thermodynamic limit is obtained by projecting onto the eigenspace associated with the eigenvalue $\lambda_{\rm max}$. 
}

\added{
Using the vanishing condition \eqref{eq:vanishing_conf} and the combinatorial factors associated with configurations selected by the boundary condition, the leading contribution consists of the MPS tensors ${^{t_p}\!}\vec{A}_{a'}^{*}\vec{A}_a = T_{\bm{1}}$ and ${^{t_p}\!}\vec{B}_{a'}^{*}\vec{B}_a$
in both the numerator and denominator. 
}

\added{
As a result, the system-size dependence and the $O(1)$ factor $\alpha_{\rm ex}^{(1)}$ cancel between the numerator and denominator, yielding the same asymptotic string correlator as the ground state:
\begin{align}
    &O^{(N,\pm)}_{\rm string}(\ell_1,\ell_2;S^z,g_z) \nonumber \\
    &= \lambda_{\rm max}^{-2} \langle \lambda_{\rm max}| T_{S^z} \sigma^z_a |\lambda_{\rm max} \rangle \langle \lambda_{\rm max}| \sigma^z_a T_{S^z} |\lambda_{\rm max} \rangle + O(L^{-1}) \nonumber \\
    &= -\frac{4}{9} + O(L^{-1}), 
    \label{eq:SOscaling}
\end{align}
as long as the number of bimagnons $N$ remains fixed in the dilute regime $N/L \to 0$.} \add{We note that the combinatorial hierarchy between contributions associated with different particle numbers remains well separated for $N^2 \ll L$; the fixed-$N$ limit considered here automatically satisfies this condition. } 
Based on \eqref{eq:SOscaling}, the string order parameter in the $N$-bimagnon state is then defined by
\begin{align}
    O^{(N,\pm)}_{\rm string}
    = \lim_{|\ell_2-\ell_1|\to\infty} \lim_{L\to\infty} \, O^{(N,\pm)}_{\rm string}(\ell_1,\ell_2;S^z,g_z).
\end{align}
\added{
The details of the evaluation can be found in Appendix \ref{ap:SO}, where we also clarify why this approach breaks down for states in the finite-density regime or superpositions of different particle-number sectors. 
}

\added{This demonstrates that the symmetry-associated topological signature of the ground state, as captured by the string order parameter, persists throughout the bimagnon scar tower in the dilute regime with fixed quasiparticle number. The present analysis does not directly extend to the finite-density regime or to superpositions of different particle-number sectors, as discussed in Appendix~\ref{ap:SO}. 
Finally, we note that ordinary two-point correlations in the bimagnon scar tower decay exponentially with distance, in contrast to the long-range string order. }

\subsubsection{Numerical confirmation}

To verify the analytical predictions for both the ground state and the bimagnon QMBS, we numerically evaluate the string correlator \eqref{eq:string_order_def} and the ordinary two-point correlation 
\begin{align}
    O_{zz}(\ell_1,\ell_2)
    &= \frac{\langle \Psi| S_{\ell_1}^z S_{\ell_2}^z |\Psi\rangle}{\langle \Psi|\Psi\rangle}
     - \frac{\langle \Psi| S_{\ell_1}^z |\Psi\rangle}{\langle \Psi|\Psi\rangle}
        \frac{\langle \Psi| S_{\ell_2}^z |\Psi\rangle}{\langle \Psi|\Psi\rangle},  \label{eq:AKLT_spin_correlation}
\end{align}
in finite-size AKLT chains under periodic boundary conditions. The subtraction of the second term is required since the bimagnon QMBS generally have nonvanishing magnetization, $\langle S^z_\ell\rangle \neq 0$. 

We diagonalize the AKLT Hamiltonian \eqref{eq:AKLT} by exact diagonalization within each symmetry sector labeled by \added{the total magnetization $S_{\mathrm{tot}}^z = \sum_i S_i^z$, total momentum $k (=0, \pi)$, and the inversion parity $I (=\pm 1)$.}
The ground state is nondegenerate with eigenenergy $E_0 = -2L/3$, while the bimagnon QMBS form a tower of excitations with total spin $S = 2N$ and eigenenergies $E_N = -2L/3 + 4N$ ($N = 1,2,\dots$)~\cite{Moudgalya2020b2, Lin2020, Mark2020, Matsui2024}.  
For convenience in presenting the numerical results, we label these spin–$2N$ bimagnon QMBS as $S_{2N}$.

\begin{figure}
    \centering
    \includegraphics[width=0.9\linewidth]{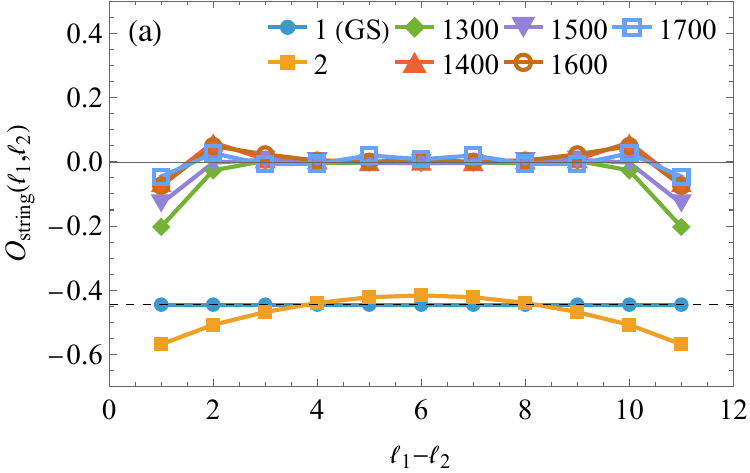}
    \includegraphics[width=0.9\linewidth]{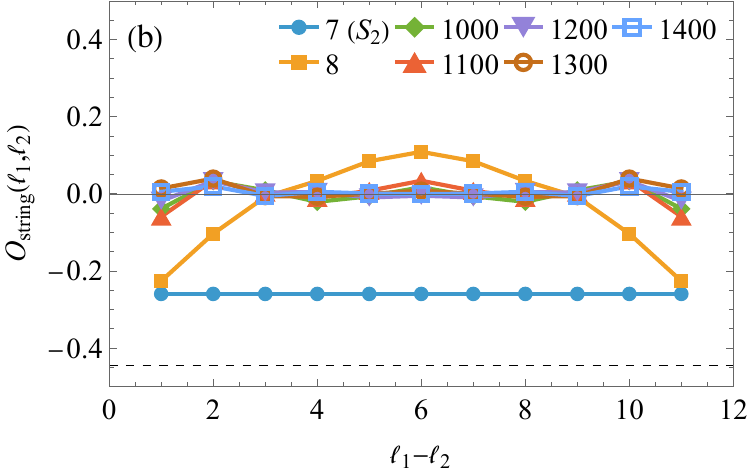}
    \includegraphics[width=0.9\linewidth]{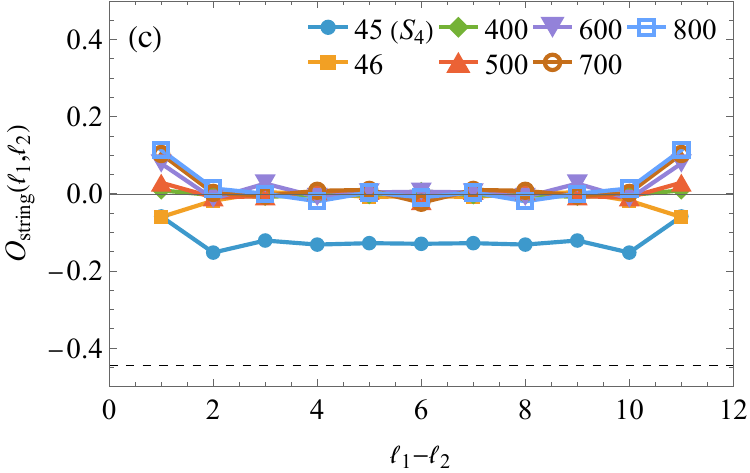}
    \includegraphics[width=0.9\linewidth]{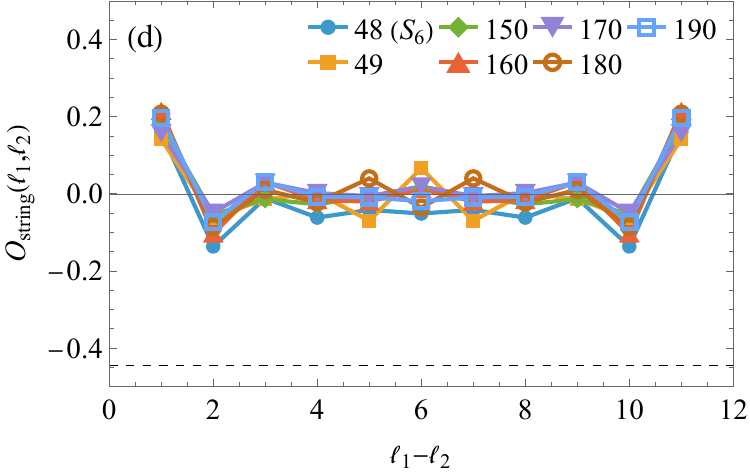}
    \caption{
    String order $O_{\rm string}(\ell_1, \ell_2;S^z,g_z)$ [Eq.~(\ref{eq:string_order_def})] for several eigenstates of the AKLT model with
    \added{(a) $L=12$, $S_{\rm tot}^z=0$, $k=0$, $I=1$,
    (b) $L=12$, $S_{\rm tot}^z=2$, $k=\pi$, $I=-1$,
    (c) $L=12$, $S_{\rm tot}^z=4$, $k=0$, $I=1$, and
    (d) $L=12$, $S_{\rm tot}^z=6$, $k=\pi$, $I=-1$.}
    The labels show the numbering of the eigenstates counted from the lowest energy state in the corresponding symmetry sector.
    GS, $S_2$, $S_4$, and $S_6$ refer to the ground state, spin 2, 4, and 6 bimagnon states, respectively.
    The dashed lines indicate the analytical value $-\frac{4}{9}$ in the large system-size and long-distance limits.
    \add{The Hilbert space dimension is 3179 (a), 2456 (b), 1244 (c), and 347 (d).}
    }
    \label{fig: string order}
\end{figure}

\begin{figure}
    \centering
    \includegraphics[width=0.9\linewidth]{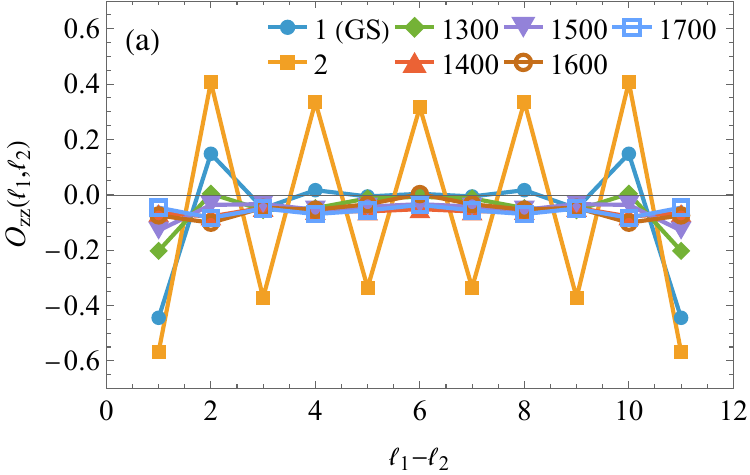}
    \includegraphics[width=0.9\linewidth]{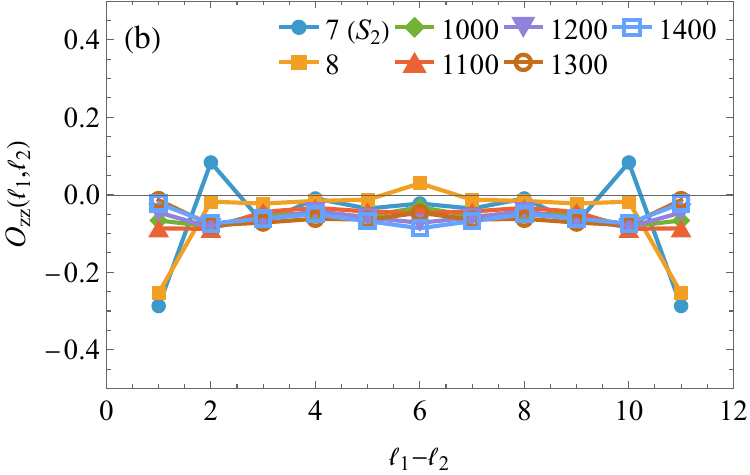}
    \includegraphics[width=0.9\linewidth]{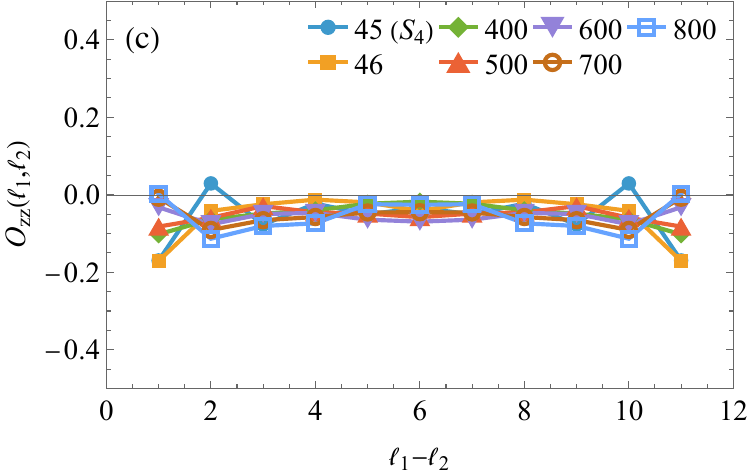}
    \includegraphics[width=0.9\linewidth]{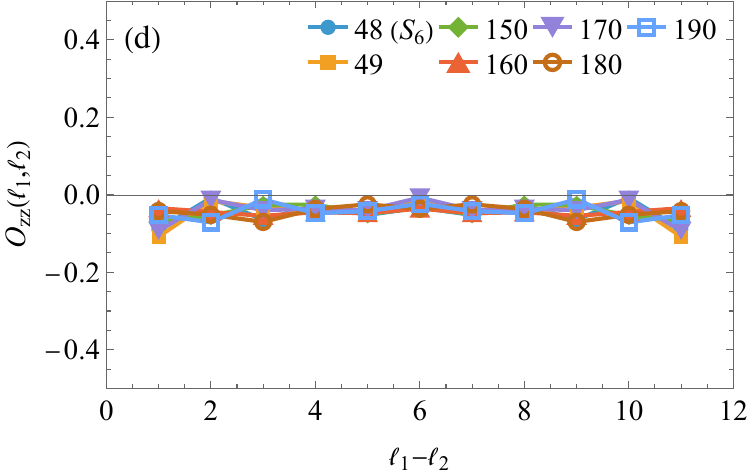}
    \caption{
    Spin-spin correlation function $O_{zz}(\ell_1, \ell_2)$ [Eq.~(\ref{eq:AKLT_spin_correlation})] for several eigenstates of the AKLT model with
    \added{(a) $L=12$, $S_{\rm tot}^z=0$, $k=0$, $I=1$,
    (b) $L=12$, $S_{\rm tot}^z=2$, $k=\pi$, $I=-1$,
    (c) $L=12$, $S_{\rm tot}^z=4$, $k=0$, $I=1$, and
    (d) $L=12$, $S_{\rm tot}^z=6$, $k=\pi$, $I=-1$.}
    The labels show the numbering of the eigenstates counted from the lowest energy state in the corresponding symmetry sector.
    GS, $S_2$, $S_4$, and $S_6$ refer to the ground state, spin 2, 4, and 6 bimagnon states, respectively.
    \add{The Hilbert space dimension is 3179 (a), 2456 (b), 1244 (c), and 347 (d).}
    }
    \label{fig: Sz correlation}
\end{figure}

\begin{figure}
    \centering
    \includegraphics[width=0.9\linewidth]{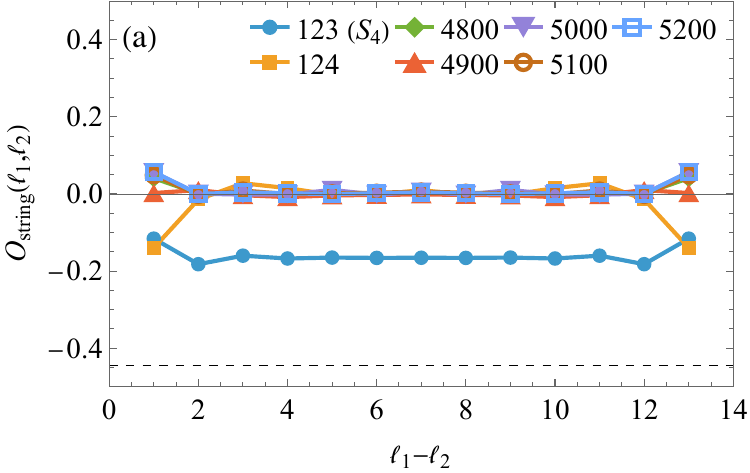}
    \includegraphics[width=0.9\linewidth]{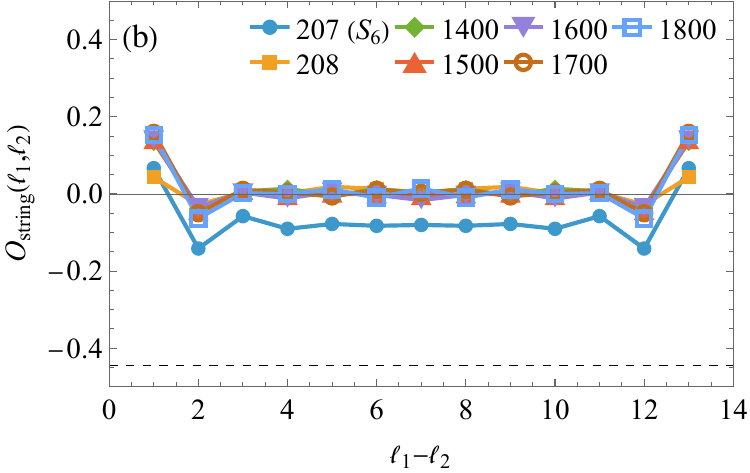}
    \caption{
    String order $O_{\rm string}(\ell_1, \ell_2;S^z,g_z)$ [Eq.~(\ref{eq:string_order_def})] for several eigenstates of the AKLT model with
    \added{(a) $L=14$, $S_{\rm tot}^z=4$, $k=0$, $I=1$ and
    (b) $L=14$, $S_{\rm tot}^z=6$, $k=\pi$, $I=-1$.}
    The labels show the numbering of the eigenstates counted from the lowest energy state in the corresponding symmetry sector.
    $S_4$ and $S_6$ refer to the spin 4 and 6 bimagnon states, respectively.
    The dashed lines indicate the analytical value $-\frac{4}{9}$ in the large system-size and long-distance limits.
    \add{The Hilbert space dimension is 9840 (a) and 3344 (b).}
    }
    \label{fig: string order L=14}
\end{figure}

\begin{figure}
    \centering
    \includegraphics[width=0.9\linewidth]{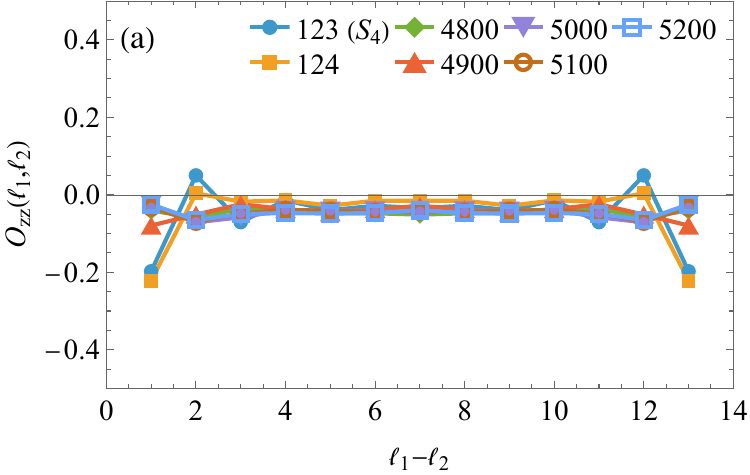}
    \includegraphics[width=0.9\linewidth]{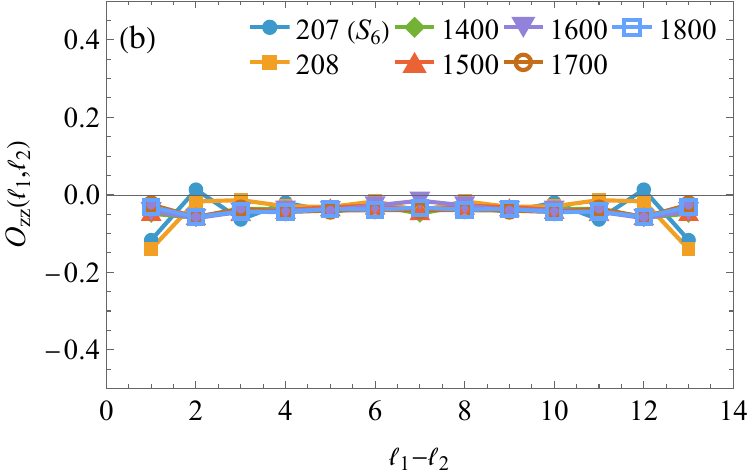}
    \caption{
    Spin-spin correlation function $O_{zz}(\ell_1, \ell_2)$ [Eq.~(\ref{eq:AKLT_spin_correlation})] for several eigenstates of the AKLT model with
    \added{(a) $L=14$, $S_{\rm tot}^z=4$, $k=0$, $I=1$ and
    (b) $L=14$, $S_{\rm tot}^z=6$, $k=\pi$, $I=-1$.}
    The labels show the numbering of the eigenstates counted from the lowest energy state in the corresponding symmetry sector.
    $S_4$ and $S_6$ refer to the spin 4 and 6 bimagnon states, respectively.
    \add{The Hilbert space dimension is 9840 (a) and 3344 (b).}
    }
    \label{fig: Sz correlation L=14}
\end{figure}

For each eigenstate $|\Psi\rangle$, whether scar or non-scar, we compute the string correlator defined in \eqref{eq:string_order_def} (Figs.~\ref{fig: string order} and \ref{fig: string order L=14} for $L=12$ and $14$). We also evaluate the ordinary two-point correlator \eqref{eq:AKLT_spin_correlation} (Figs.~\ref{fig: Sz correlation} and \ref{fig: Sz correlation L=14} for $L=12$ and $14$), for comparison of their long-distance behavior. 

Figure~\ref{fig: string order} shows the string correlator $O_{\rm string}(\ell_1,\ell_2)$ for several eigenstates \added{at $L=12$ with
(a) $S_{\rm tot}^z=0$,
(b) $S_{\rm tot}^z=2$,
(c) $S_{\rm tot}^z=4$, and
(d) $S_{\rm tot}^z=6$.}
We classify the eigenstates according to the quantum numbers, $L$, $S_{\rm tot}^z$, $k$, and $I$.
Each eigenstate is labeled by the numbering counted from the lowest energy state in the corresponding symmetry sector.

\begin{figure}
    \centering
    \includegraphics[width=0.9\linewidth]{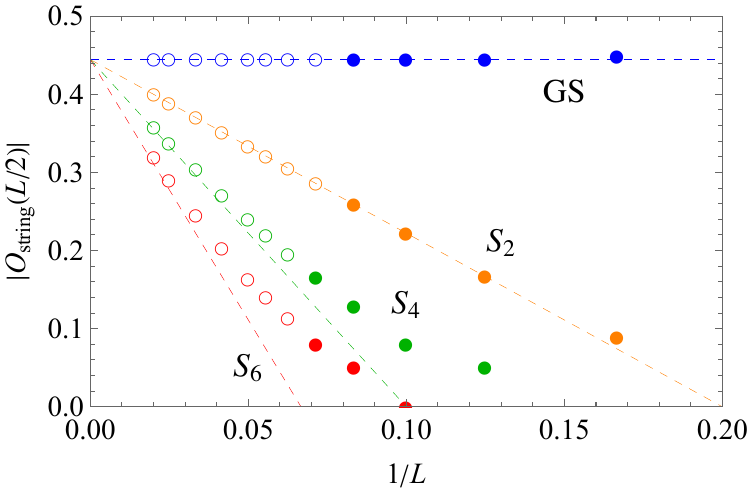}
    \caption{\add{Scaling plot of the string order $|O_{\rm string}(\ell_1,\ell_1+L/2)|$ for the ground states (GS) and bimagnon QMBS states ($S_2, S_4, S_6$). 
    The dashed lines show the asymptotic form of the string order, $4/9$ and $4/9-20N/(9L)$, for the GS and bimagnon states $S_{2N}$, respectively. The filled (open) circles represent the results obtained by exact diagonalization (MPS calculations).}
    %The black dashed line shows the value in the thermodynamic limit, $4/9$, while the red dashed line shows the asymptotic form, $4/9-20/(9L)$, of the string order for the bimagnon state $S_2$.}
    }
    \label{fig:scaling plot}
\end{figure}

\new{The ground state exhibits a constant value which is very close to the thermodynamic limit, $O_{\rm string}=-4/9$, consistent with the analytical result \eqref{eq:string_gs}.}
Remarkably, each bimagnon QMBS ($S_{2N}$) also maintains a nearly constant string order over long distances, confirming the persistence of the nonlocal order in these QMBS.
While the magnitude of the string order ($|O_{\rm string}|$) is much smaller than $4/9$ for finite $L$, this suppression is attributed to finite-size effects, and is expected to vanish in the large-$L$ limit. In fact, if we look at the results for $L=14$ and $S_{\rm tot}^z=4, 6$ (Fig.~\ref{fig: string order L=14}), the string order of the bimagnon states $S_{2N}$ tends to be enhanced as the system size $L$ grows with $N$ fixed.
In Fig.~\ref{fig:scaling plot}, we show the scaling plot of $|O_{\rm string}|$ measured with the interval $L/2$ for the ground states (GS) and bimagnon QMBS states ($S_2, S_4, S_6$). 
\add{The filled and open circles represent the results obtained from exact diagonalization and MPS calculations, respectively. We have confirmed that both of the methods give the same results for small $L$.

We can clearly see that the magnitude of the string order for each QMBS state approaches $4/9$ in the thermodynamic limit with $N$ fixed. 
To characterize the finite-size corrections, we explicitly evaluate the string order using the exact MPS representation of the bimagnon states~\cite{Moudgalya:2018PhRvB..98w5156M}. For fixed $N$ with $0\leq N\leq5$, we obtain
\begin{align} \label{eq:SOsubleading}
    O^{(N,\pm)}_{\rm string}(\ell_1,\ell_2;S^z,g_z)
    = -\frac{4}{9} + \frac{20}{9}\frac{N}{L} + O(L^{-2}). 
\end{align}
The numerical results for $S_{2N}$ agree very well with the asymptotic form \eqref{eq:SOsubleading}, shown by the dashed lines in Fig.~\ref{fig:scaling plot}. 
}

Generic eigenstates outside the scar tower, on the other hand, display a rapidly decaying string order.
In Figs.~\ref{fig: string order} and \ref{fig: string order L=14}, we plot $O_{\rm string}$ for several eigenstates taken from the middle of the spectrum in the corresponding symmetry sector. For reference, the Hilbert space dimension is given by 3179 (Fig.~\ref{fig: string order}(a)), 2456 (b), 1244 (c), 347 (d), 9840 (Fig.~\ref{fig: string order L=14}(a)), and 3344 (b), respectively.
\add{We also plot $O_{\rm string}$ for the neighboring eigenstate next to the ground state and QMBS states in Fig.~\ref{fig: string order} and \ref{fig: string order L=14}.}
We can see that the string order is generically suppressed outside of the QMBS subspace \add{(except for those close to the ground state)}, indicating the absence of long-range nonlocal order.

To highlight the distinction between the string order and ordinary long-range order, Fig.~\ref{fig: Sz correlation} ($L=12$) and Fig.~\ref{fig: Sz correlation L=14} ($L=14$) show the two-point spin-spin correlator $O_{zz}$ (\ref{eq:AKLT_spin_correlation}) for the same eigenstates as in Fig.~\ref{fig: string order} and Fig.~\ref{fig: string order L=14}, respectively.
As expected, for all the eigenstates, including the ground state and the scar states, the ordinary two-point correlators $O_{zz}$ decay rapidly as the distance $|\ell_2 - \ell_1|$ increases. We do not clearly observe an enhancement of $O_{zz}$ for the bimagnon QMBS states $S_{2N}$ in the large-$L$ limit with $N$ fixed. 

Thus, we conclude that the persistence of nonlocal string order, rather than any local correlation, is the hallmark of the topological structure shared by the ground state and the scar subspace. These numerical results verify that the nonlocal string order persists throughout the entire bimagnon scar subspace, supporting our analytical findings. 

\section{Realizations beyond the AKLT model}
\label{sec:robustness}

The discussions so far have established that the quasiparticle tower of QMBS, when constructed on a nontrivial SPT ground state, naturally forms \soften{a symmetry-invariant scar subspace composed of small-bond MPS eigenstates, within which symmetry-associated topological properties can be consistently defined. In the present construction, this structure is realized through the rSGA}.  
\soften{We now demonstrate that analogous structures arise in a broader class of models admitting an rSGA, beyond the AKLT model.}

\added{As representative examples, we construct a parametrized spin-$1$ Hamiltonian that preserves inversion, time-reversal, and on-site $\mathbb{Z}_2 \times \mathbb{Z}_2$ symmetries, together with the rSGA structure generated by the bimagnon creation operators. For this model, we verify that the symmetry-associated topological properties manifest through the bond-space representations and their observable signatures. 
In addition, we demonstrate that similar topological signatures arise in higher-spin AKLT models~\cite{Zang2010, Jiang2010,Moudgalya:2018PhRvB..98w5156M}, where the generalized VBS ground states serve as reference states for scar subspaces. }

\added{As a common setting, we consider a class of Hamiltonians that preserve inversion, time-reversal, and on-site $\mathbb{Z}_2 \times \mathbb{Z}_2$ symmetries, and host a bimagnon scar subspace exhibiting nontrivial symmetry-associated topological properties. 
This structure is realized on a reference state given by a symmetry-invariant MPS with finite bond dimension, associated with nontrivial topology: 
\begin{align} 
    |\Phi_0 \rangle = \sum_{\substack{m_1,\dots,m_L \\ \in \{0,1,\dots,2s\}}}
    &{\rm tr}\,(A_{m_1}A_{m_2}\cdots A_{m_L}) 
     |m_1 m_2 \dots m_L \rangle. 
\end{align}
Two species of QMBS towers are constructed on this MPS reference state by repeated application of quasiparticle creation operators $Q^{\pm}$ \eqref{eq:rSGA_AKLT}:
\begin{align} \label{eq:bimagnon_pert}
    |\Phi_N^{\pm}\rangle = (Q^{\pm})^N |\Phi_0\rangle,
\end{align}
which together span a subspace $W_{\mathrm{pert}} = \mathrm{span}\{|\Phi^+_N\rangle, |\Phi^-_N\rangle\}_N$ that is invariant under inversion, time-reversal, and on-site $\mathbb{Z}_2 \times \mathbb{Z}_2$ symmetries.
}

\subsection{Deformations of the AKLT model}

\added{As a concrete realization of this structure, we construct a two-parameter family of Hamiltonians that admit such a bimagnon scar subspace, with the AKLT point recovered at special parameter values. }
\added{We consider a reference state given by a symmetry-invariant MPS with a two-dimensional bond space, }
\begin{align} \label{eq:MPS_gs_pert}
    |\Phi_0 \rangle = \sum_{\substack{m_1,\dots,m_L \\ \in \{0,1,2\}}} 
    \mathrm{tr}\,(A_{m_1}A_{m_2}\cdots A_{m_L}) |m_1 m_2 \dots m_L \rangle, 
\end{align}
\added{associated with nontrivial topology, where the MPS tensors are given by}
\begin{align} 
    &A_0 = \frac{\gamma}{\sqrt{1+\gamma^2}}\sigma_a^+, \quad
    A_1 = -\frac{1}{\sqrt{1+\gamma^2}}\sigma_a^z, \nonumber \\
    &A_2 = -\frac{\gamma}{\sqrt{1+\gamma^2}}\sigma_a^-, \label{eq:MPS_gs_pert2}
\end{align}
with $\gamma \in \mathbb{R} \setminus \{0\}$. For any $\gamma \neq 0$, the resulting MPS is invariant under inversion, time-reversal, and on-site $\mathbb{Z}_2 \times \mathbb{Z}_2$ symmetries. The AKLT state is recovered for $\gamma=\sqrt{2}$.

\added{Bimagnon QMBS of the form \eqref{eq:bimagnon_pert}, created by $Q^{\pm}$ \eqref{eq:rSGA_AKLT}, span a subspace $W_{\mathrm{pert}} = \mathrm{span}\{|\Phi^+_N\rangle, |\Phi^-_N\rangle\}_N$ that is invariant under inversion, time-reversal, and on-site $\mathbb{Z}_2 \times \mathbb{Z}_2$ symmetries. }
A translationally invariant Hamiltonian satisfying the rSGA relation~\eqref{eq:rSGA_AKLT} within the subspace \added{$W_{\rm pert}$} takes the form
\begin{align} \label{eq:perturbed}
    H_{\mathrm{pert}} = \sum_{j=1}^L h_{j,j+1},
\end{align}
where the local Hamiltonian is given in the local basis $\{ |0\rangle,|1\rangle,|2\rangle \}$ by~\footnote{The local term can also be rewritten in terms of spin-$1$ operators, but the resulting expression is cumbersome and does not provide additional insight for the present discussion.}
\begin{align}
    h_{j,j+1} &= 
    \mathcal{E}(|00\rangle\langle00| + |22\rangle\langle22|) \nonumber \\
    &+ \frac{\mathcal{E}}{2} \sum_{a=0,2} 
    (|a1\rangle\langle a1| + |1a\rangle\langle1a|
    + |a1\rangle\langle1a| + |1a\rangle\langle a1|) \nonumber \\
    &+ \alpha (|02\rangle\langle02| + |20\rangle\langle20|
    + |02\rangle\langle20| + |20\rangle\langle02|) \nonumber \\
    &+ \alpha\gamma (|02\rangle\langle11| + |11\rangle\langle02|
    + |20\rangle\langle11| + |11\rangle\langle20|) \nonumber \\
    &+ \alpha\gamma^2 |11\rangle\langle11| .
    \label{eq:pert_H}
\end{align}
The parametrized Hamiltonian \eqref{eq:pert_H} includes the AKLT model at $\mathcal{E}=1$, $\alpha=1/6$, and $\gamma=\sqrt{2}$, up to an additive constant and an overall multiplicative factor. 

The terms proportional to $\mathcal{E}$ in the first two lines of \eqref{eq:pert_H} fix the level spacing of the equally spaced spectrum in the scar subspace $W_{\rm pert}$ through the rSGA relation. Consequently, at fixed spacing $2\mathcal{E}$, \added{the deformed Hamiltonian}, preserving the on-site $\mathbb{Z}_2\times\mathbb{Z}_2$, inversion, and time-reversal symmetries together with the rSGA structure, is fully characterized by two real parameters, $\alpha$ and $\gamma\neq 0$.  
In contrast, the remaining terms in \eqref{eq:pert_H} commute with the bimagnon creation operators $Q^\pm$ and therefore act as multiples of the identity within $W_{\rm pert}$, without shifting the eigenenergies of the scar manifold. 
\added{Although the parameter $\gamma$ enters the MPS ground state, it does not alter the underlying topology, but only affects the quantitative values of observables, as discussed in the following subsections. }

\subsubsection{Topological properties of the deformed scar subspace}

Since the MPS ground state \eqref{eq:MPS_gs_pert}, \eqref{eq:MPS_gs_pert2} inherits all symmetries of the perturbed Hamiltonian~\eqref{eq:pert_H}---inversion, time-reversal, and the on-site $\mathbb{Z}_2\times\mathbb{Z}_2$ symmetry---and remains injective throughout the full two-parameter family (for $\gamma \neq 0$ and $\alpha \in \mathbb{R}$), its symmetry-associated topological properties are completely determined by the induced symmetry actions on the bond space.
Consequently, the ground-state tensors obey the transformation law
\begin{align}\label{eq:bond_transformation_pert_ground}
    &\mathcal{S}:~\vec{A}\;\mapsto\; e^{i\theta_{\mathcal{S}}} \,V_{\mathcal{S}}\,\vec{A}\,V_{\mathcal{S}}^{\dagger}, \\
    &\mathcal{S}\in\{\mathcal{P},\mathcal{T},g_z,g_x\}, \nonumber
\end{align}
where $g_x$ and $g_z$ generate $G = \mathbb{Z}_2 \times \mathbb{Z}_2$. The corresponding virtual-space operators are given by 
\begin{align}
    &V_{\mathcal{P}} = i\sigma^{y},\nonumber\\
    &V_{\mathcal{T}} = i\sigma^{y},\nonumber\\
    &V_{g_z} = \sigma^{z},\quad V_{g_x} = \sigma^{x}. \label{eq:bond_generators_pert_ground}
\end{align}
These matrices define the same projective representation of the symmetry group for all $\gamma \neq 0$ and $\alpha \in \mathbb{R}$, including the AKLT point. Therefore, the ground state remains in the same cohomology class as the AKLT ground state. 

\added{As the MPO for the bimagnon-creation operator $Q^{\pm}$ is identical to that of the AKLT model, the symmetry-associated topological properties of the bimagnon scar subspace are entirely determined by those of the ground-state MPS. Consequently, the scar subspace $W_{\rm pert}$ retains the same symmetry fractionalization structure as in the AKLT case. }

This shows that the full SPT-like structure of the bimagnon scar tower \added{is preserved throughout} two-parameter deformation \eqref{eq:pert_H}.

\subsubsection{String order in the deformed scar subspace}

\added{We next examine the behavior of the long-range string order parameter as a representative observable associated with the underlying symmetry-associated topological properties. 
In close analogy with the AKLT case, the string correlator for the $N$-bimagnon states $|\Phi^{\pm}_N \rangle$ is evaluated as}
\begin{align}
    \widetilde{O}_{\mathrm{string}}^{(N,\pm)}(\ell_1,\ell_2;S^z,g_z;\gamma)
    &= \frac{\langle \Phi_N^{\pm}| S_{\ell_1}^z
    \cdot
    \prod_{\ell = \ell_1+1}^{\ell_2-1} R_{\ell}^z
    \cdot
    S_{\ell_2}^z
    |\Phi_N^{\pm}\rangle}
    {\langle \Phi_N^{\pm}|\Phi_N^{\pm}\rangle} \nonumber \\
    &= -\left(\frac{\gamma^2}{1+\gamma^2}\right)^{\!2} + O(L^{-1}). 
\end{align}
\added{Thus, in the dilute regime, the string order parameter $\widetilde{O}_{\mathrm{string}}^{(N,\pm)}(\gamma) = \lim_{|\ell_2 - \ell_1| \to \infty} \lim_{L \to \infty} \widetilde{O}_{\mathrm{string}}^{(N,\pm)}(\ell_1,\ell_2;S^z,g_z;\gamma)$ depends only on the MPS parameter $\gamma$. }

\added{This demonstrates that the same symmetry-associated topological structure is realized throughout the scar subspace in the dilute regime. }

\subsection{Higher-spin generalization of the AKLT model}

We next consider higher-spin generalizations. 
For integer spin $s$, a generalized valence-bond-solid (VBS) state forms $2s$ singlet bonds between nearest neighbors~\cite{Moudgalya2018}, which admits an MPS representation of the form
\begin{align} \label{eq:generalizedVBS}
    |\Phi_0 \rangle = \sum_{\substack{m_1,\dots,m_L \\ \in \{0,1,\dots,2s\}}} {\rm tr} (A_{m_1} \cdots A_{m_L}) |m_1 \cdots m_L \rangle.
\end{align}
The local tensors $A_m$ are constructed from Clebsch--Gordan coefficients, corresponding to the symmetrization of $2s$ virtual spin-$1/2$ degrees of freedom into the physical spin-$s$ representation~\cite{Moudgalya:2018PhRvB..98w5156M}. 
This MPS provides a natural higher-spin setting for analyzing symmetry-associated topological properties under inversion, time-reversal, and on-site $\mathbb{Z}_2 \times \mathbb{Z}_2$ symmetries. In particular, when $s$ is odd, the state \eqref{eq:generalizedVBS} exhibits nontrivial symmetry-associated topological properties~\cite{Oshikawa:1992JPCM....4.7469O,Pollmann2012}. 

The generalized VBS state \eqref{eq:generalizedVBS} arises as the unique gapped ground state of the integer spin-$s$ AKLT Hamiltonian,
\begin{align}
    H = \sum_{j=1}^L \sum_{J=s+1}^{2s} \alpha_J P^{(J,s)}_{j,j+1},
\end{align}
where $\alpha_J > 0$ for all $J$~\cite{Zang2010, Jiang2010}. $P^{(J,s)}_{j,j+1}$ denotes the projector onto the total-spin-$J$ sector of two neighboring spin-$s$ sites.

\subsubsection{Symmetry-invariant scar subspace in the higher-spin AKLT model}

The spin-$s$ AKLT model admits exact quasiparticle excitations created by~\cite{Moudgalya2018, Moudgalya:2018PhRvB..98w5156M}
\begin{align} \label{eq:AKLTs_creation}
    Q^{\pm} = \sum_j (-1)^j (S^{\pm}_j)^{2s}. 
\end{align}
These operators locally break a valence bond and create isolated excitations, in direct analogy with the spin-$1$ case. 

Due to the constraint structure of the VBS state, the action of $Q^{\pm}$ produces states in which excitations cannot overlap or occupy adjacent sites, leading to an effective hard-core behavior. 
As a result, repeated application of $Q^{\pm}$ generates a tower of states
\begin{align}
    |\Phi_n^{\pm}\rangle = (Q^{\pm})^n |\Phi_0 \rangle,
\end{align}
which form an invariant subspace
\begin{align}
    W = {\rm span}\{ |\Phi_n^+\rangle,\, |\Phi_n^-\rangle \}_n.
\end{align}

Within this subspace, the Hamiltonian satisfies the restricted spectrum-generating algebra (rSGA) relation
\begin{align}
    ([H,\,Q^{\pm}] - 2\mathcal{E} Q^{\pm}) \big|_W = 0,
\end{align}
where $\mathcal{E}$ is a constant that sets the energy spacing of the tower. 
This subspace is invariant under inversion, time-reversal, and the on-site $\mathbb{Z}_2 \times \mathbb{Z}_2$ symmetries, as discussed in Sec.~\ref{subsec:inv_subspace}.

\subsubsection{Topological properties of the higher-spin scar subspace}

Since the generalized VBS state \eqref{eq:generalizedVBS} exhibits nontrivial symmetry-associated topological properties under inversion, time-reversal, and on-site $\mathbb{Z}_2 \times \mathbb{Z}_2$ symmetries~\cite{Oshikawa:1992JPCM....4.7469O,Pollmann2012}, these symmetry actions 
\begin{align}
    &\mathcal{S}:~ \vec{A} \mapsto 
    e^{i\theta_{\mathcal{S}}}\,
    V_{\mathcal{S}}\, \vec{A}\, V_{\mathcal{S}}^{\dagger},
    \\
    &\mathcal{S}\in\{\mathcal{P},\,\mathcal{T},\,g_x,\,g_z\}
\end{align}
are represented projectively on the bond space, where $g_x$ and $g_z$ generate the on-site $\mathbb{Z}_2 \times \mathbb{Z}_2$ symmetry. 
The matrices $V_{\mathcal{S}}$ form a projective representation on the $(s+1)$-dimensional bond space, belonging to the same cohomology class as in the spin-$1$ AKLT case for odd integer $s$~\cite{Pollmann2010, Pollmann2012}. 

As a translationally invariant local excitation, the creation operator \eqref{eq:AKLTs_creation} admits the MPO expression \eqref{eq:MPO_local}
\begin{align}
    &Q^{\pm} = {_b}\langle \uparrow\!| M_{\pm} \otimes_p \cdots \otimes_p M_{\pm} |\!\downarrow \rangle_b, \nonumber \\
    &M_{\pm} = \begin{pmatrix} -\bm{1} & -(S^{\pm})^{2s} \\ 0 & \bm{1} \end{pmatrix}_b, 
\end{align}
by which the symmetry properties of these excitations can be analyzed. The MPO tensors transform projectively under inversion
\begin{align}
    \mathcal{P}:~M^{\pm} \mapsto V_{\mathcal{P}} M^{\pm} V_{\mathcal{P}}^\dagger
\end{align}
with $V_{\mathcal{P}} = i\sigma^y$, while they transform with trivial bond representations under time-reversal and $g_z,g_x \in \mathbb{Z}_2 \times \mathbb{Z}_2$ transformations, 
\begin{align}
    &\mathcal{T}:~ M^{\pm}
    \mapsto M^{\mp}, \\
    &g_z:~M_{\pm} \mapsto M_{\pm}, \nonumber \\
    &g_x:~M_{\pm} \mapsto M_{\mp}.  
\end{align}
Hence, the resulting scar subspace faithfully inherits the symmetry-associated topological structure of the underlying SPT ground state under time-reversal and on-site $\mathbb{Z}_2 \times \mathbb{Z}_2$ symmetries, while the inversion symmetry systematically alternates the topology between trivial and nontrivial sectors depending on the quasiparticle number $N$, in direct analogy with the spin-$1$ AKLT model discussed in Sec.~\ref{subsec:bond_rep}. 

Building on the above observations, we conclude that the symmetry-based construction of SPT-like scar subspaces extends beyond the spin-$1$ AKLT model to a broader class of VBS-type systems. 
These systems provide a natural setting in which 
(i) an SPT ground state, 
(ii) a symmetry-invariant subspace composed of finite-bond-dimension MPS states, and 
(iii) consistent symmetry actions on the bond space, 
are simultaneously realized. 
The construction is therefore not specific to the AKLT model, but instead provides a general framework for realizing symmetry-associated topological structures in scar subspaces.

\section{Conclusion}
\label{sec:conclusion}

In this work, we have proposed that \soften{symmetry-associated topological properties} of an SPT ground state can persist in a dynamically isolated subspace of QMBS, provided that the subspace is closed under the \added{relevant} symmetry (inversion, time reversal, or on-site symmetry). Although individual QMBS need not be symmetric on their own, we showed that the subspace spanned by the QMBS retains the full \added{symmetry structure of the ground state, inherited from the Hamiltonian}. Based on this observation, we introduced the notion of \soften{an \emph{SPT-like scar subspace}}, defined as a symmetry-invariant scar subspace in which all QMBS connected by the protecting symmetry transformations share the same cohomology class of the bond space symmetry action, and therefore exhibit well-defined topological properties. 

As a concrete example, we analyzed the Affleck–Kennedy–Lieb–Tasaki (AKLT) model.  
By examining the symmetry representations on the bond space, we \added{showed that} the bimagnon scar subspace \added{either directly} inherits \added{or modifies} the topological characteristics of the ground state, \added{depending on the} symmetries. 
\added{We further showed that this subspace exhibits characteristic signatures of SPT systems, including nontrivial expectation values of symmetry-twist operators and a non-zero long-range string order in the dilute regime, which distinguish the QMBS from generic excited states. }
\new{We also showed that this symmetry-based characterization remains valid under deformations and higher-spin generalizations of the AKLT model that preserve the essential structural ingredients. This indicates that SPT-like scar subspaces arise more broadly in VBS-type systems. }

\added{The present results point toward a broader perspective in which symmetry-invariant subspaces, rather than individual eigenstates, provide a natural setting for discussing topological properties in excited states.  
At the same time, several important questions remain open, including whether similar structures can arise in scar subspaces built on non-solvable ground states. }
\added{Addressing these questions will be essential for understanding the mechanisms that stabilize well-defined topological properties in nonthermal subspaces of quantum many-body systems. }
We leave these challenging directions for future work.

\section{Acknowledgements}
\added{We thank H. Katsura for suggesting the evaluation of the string order parameter for a coherent state. }
C.M. acknowledges support by JSPS KAKENHI (Grant No.~JP23K03244) and UTokyo Global Activity Support Program for Young Researchers. 
N.T. acknowledges support by JST FOREST (Grant No.~JPMJFR2131) and JSPS KAKENHI (Grant Nos.~JP24H00191, JP25H01246, JP25H01251).

\appendix
%%%%%%%%%%%%%%%%%%%%%%%%%%%%%%%%%%%%%%%%%%%%%%%%%%
%%%%%%%%%%%%%%%%%%%%%%%%%%%%%%%%%%%%%%%%%%%%%%%%%%
%%%%%%%%%%%%%%%%%%%%%%%%%%%%%%%%%%%%%%%%%%%%%%%%%%
\section{A Lemma on Group Cohomology}\label{ap:Lemma}

In this appendix we establish a simple but useful observation: 
when two MPS are related by the action of a symmetry element of the protecting group, the associated bond-space symmetry representations necessarily belong to the same cohomology class. This ensures that symmetry-connected QMBS carry the same projective invariant, as used in the main text.

Consider a projective representation $V: G \to GL(N)$ of a finite group $G$, characterized by a well-defined cohomology class $\omega \in H^2(G,U(1))$.
The representation matrices satisfy
\begin{align}
  \label{eq:RepPropertyApp}
  V_g V_h = \omega(g,h)\, V_{gh},
  \qquad
  V_e = \bm{1}.
\end{align}
A change of basis by any invertible matrix $S$ leaves the projective class invariant; that is, $SV_gS^{-1}$ obeys the same relation with the same $2$-cocycle $\omega$.

Let $k \in G$ and consider the inner automorphism $\Omega(g) = k g k^{-1}$. Define a new representation
$W = V \circ \Omega$, i.e. $W_g = V_{k g k^{-1}}$.
We claim that $W$ is a projective representation with the \emph{same} cohomology class $\omega$.

Rather than working directly with $W_g$, it is convenient to perform a further basis change by $S = V_{k^{-1}}$ and consider
\begin{align}
  X_g
  = S V_{k g k^{-1}} S^{-1}
  = V_{k^{-1}} V_{k g k^{-1}} V_{k^{-1}}^{-1}.
\end{align}
Such a basis change does not modify the projective class.
Using the representation property~\eqref{eq:RepPropertyApp}, we compute
\begin{align}
  &V_{k g k^{-1}}
    = \omega(k g, k^{-1})^{-1} \omega(k,g)^{-1}
       V_k V_g V_{k^{-1}}, \\
  &X_g
    = \omega(k g, k^{-1})^{-1} \omega(k,g)^{-1} \omega(k^{-1},k)
       \, V_g.
\end{align}
Thus $X_g$ differs from $V_g$ only by a $g$-dependent overall phase, i.e. by a gauge transformation.  Consequently, $W$ and $V$ represent the same cohomology class $\omega$.

From a mathematical standpoint, this is a direct consequence of the functoriality of group cohomology.  The argument also extends to cases in which $G$ acts nontrivially on the $U(1)$ coefficient, such as when antiunitary operations (e.g.\ time-reversal) are present.

\section{String order for $N$-bimagnon QMBS} \label{ap:SO}

%\new{
In this appendix, we analytically evaluate the string order parameter for the bimagnon QMBS, including both the thermodynamic limit and the leading finite-size corrections. In particular, we derive the asymptotic value $-4/9$ in the double limit of large system size and large separation, and further determine the subleading system-size dependence. 

Let us start from the $N=1$ case. 
From the transfer-matrix expression of the norm \eqref{eq:single_bimagnon}, only two contributions remain nonvanishing under the boundary condition $K_{b’,b}$:  
\begin{widetext}
\begin{align}
    \langle \Psi_1^{\pm}|\Psi_1^{\pm} \rangle 
    &= {\rm tr}_{b',a',a,b}\,
      \Big(K_{b',b} (\mathbb{T}^{\pm\pm}_{\bm{1}})^L \Big) \nonumber \\
    &= L\, {\rm tr}_{a',a} \Big( ({^{t_p}}\!\vec{A}_{a'}^* \vec{A}_a)^{L-1} ({^{t_p}}\!\vec{B}_{a'}^* \vec{B}_a) \Big) \nonumber \\
    &\quad + 2\,{\rm Re} \sum_{1 \le j < k \le L}
    {\rm tr}_{a',a} \Big(
    ({^{t_p}}\!\vec{A}_{a'}^* \vec{A}_a)^{j-1}
    ({^{t_p}}\!\vec{A}_{a'}^* \vec{B}_a)
    ({^{t_p}}\!\vec{A}_{a'}^* \vec{A}_a)^{k-j-1}
    ({^{t_p}}\!\vec{B}_{a'}^* \vec{A}_a)
    ({^{t_p}}\!\vec{A}_{a'}^* \vec{A}_a)^{L-k}
    \Big) \nonumber \\
    &= L\, {\rm tr}_{a',a} \Big( ({^{t_p}}\!\vec{A}_{a'}^* \vec{A}_a)^{L-1} ({^{t_p}}\!\vec{B}_{a'}^* \vec{B}_a) \Big) \nonumber \\
    &\quad + 2\, {\rm Re} \sum_{r=0}^{L-2} (L-r-1)\,
    {\rm tr}_{a',a} \Big(
    ({^{t_p}}\!\vec{A}_{a'}^* \vec{B}_a)
    ({^{t_p}}\!\vec{A}_{a'}^* \vec{A}_a)^r
    ({^{t_p}}\!\vec{B}_{a'}^* \vec{A}_a)
    ({^{t_p}}\!\vec{A}_{a'}^* \vec{A}_a)^{L-r-2}
    \Big).
\end{align}
\end{widetext}
The transfer matrix $T_{\bm{1}} = {^{t_p}}\!\vec{A}_{a’}^* \vec{A}_a$ is dominated by its dominant eigenvalue $\lambda_{\rm max}=1$, while subleading eigenvalues $\pm 1/3$ are suppressed exponentially with the separation $r$. As a result, for large $r$, the insertions $^{t_p}\!\vec{A}_{a’}^* \vec{B}_a$ and $^{t_p}\!\vec{B}_{a’}^* \vec{A}_a$ are effectively projected onto the dominant eigenspace, and all well-separated configurations contribute with the same leading asymptotic weight. 

Direct computation in the $4 \times 4$ auxiliary space shows that
${^{t_p}}\!\vec{A}_{a'}^* \vec{B}_a$, ${^{t_p}}\!\vec{B}_{a'}^* \vec{A}_a$, and their products $({^{t_p}}\!\vec{A}_{a'}^* \vec{B}_a) ({^{t_p}}\!\vec{B}_{a'}^* \vec{A}_a)$ and $({^{t_p}}\!\vec{B}_{a'}^* \vec{A}_a) ({^{t_p}}\!\vec{A}_{a'}^* \vec{B}_a)$ vanish exactly within the eigenspace associated with the dominant eigenvalue $\lambda_{\rm max}$.
Therefore, the large-$L$ behavior is determined by the first term,
\begin{align}
    \langle \Psi_1^{\pm}|\Psi_1^{\pm} \rangle
    =
    L \,\lambda_{\rm max}^{L-1}
    \langle \lambda_{\rm max}| {^{t_p}}\!\vec{B}_{a'}^* \vec{B}_a |\lambda_{\rm max} \rangle
    + O(1).
\end{align}

Applying the same large-$L$ analysis to the numerator of the string correlator, we obtain the leading order expression given by 
\begin{widetext}
\begin{align}
    &O_{\rm string}^{1,\pm}(\ell_1,\ell_2;S^z,g_z) \nonumber \\
    &= 
    \frac{(L-2) \lambda_{\rm max}^{L-3} 
    (\langle \lambda_{\rm max}| {^{t_p}}\!\vec{B}_{a'}^* \vec{B}_a |\lambda_{\rm max} \rangle) 
    \langle \lambda_{\rm max}| {^{t_p}}\!\vec{A}_{a'}^* S^z \sigma^z_a \vec{A}_a |\lambda_{\rm max} \rangle 
    \langle \lambda_{\rm max}| {^{t_p}}\!\vec{A}_{a'}^* \sigma^z_a S^z \vec{A}_a |\lambda_{\rm max} \rangle}
    {L\, \lambda_{\rm max}^{L-1} 
    (\langle \lambda_{\rm max}| {^{t_p}}\!\vec{B}_{a'}^* \vec{B}_a |\lambda_{\rm max} \rangle)} + O(L^{-1}) \nonumber \\
    &= \lambda_{\rm max}^{-2} \langle \lambda_{\rm max}| {^{t_p}}\!\vec{A}_{a'}^* S^z \sigma^z_a \vec{A}_a |\lambda_{\rm max} \rangle 
    \langle \lambda_{\rm max}| {^{t_p}}\!\vec{A}_{a'}^* \sigma^z_a S^z \vec{A}_a |\lambda_{\rm max} \rangle + O(L^{-1}). 
\end{align}
\end{widetext}
This coincides with the string order parameter of the ground state. 

For a general but fixed $N$, the leading contribution is controlled by the dominant eigenvalue $\lambda_{\rm max}$ of the transfer matrix. At large separations, the transfer-matrix products effectively project the inserted tensors onto the dominant eigenspace, such that all well-separated configurations contribute with the same leading asymptotic weight. The $L$-dependence of the leading term is then determined by the number of such configurations: 
\begin{widetext}
\begin{align}
    &O_{\rm string}^{N,\pm}(\ell_1,\ell_2;S^z,g_z) \nonumber \\
    &= 
    \frac{\binom{L-2}{N}\, \lambda_{\rm max}^{L-N-2} 
    (\langle \lambda_{\rm max}| {^{t_p}}\!\vec{B}_{a'}^* \vec{B}_a |\lambda_{\rm max} \rangle)^{N} 
    \langle \lambda_{\rm max}| {^{t_p}}\!\vec{A}_{a'}^* S^z \sigma^z_a \vec{A}_a |\lambda_{\rm max} \rangle 
    \langle \lambda_{\rm max}| {^{t_p}}\!\vec{A}_{a'}^* \sigma^z_a S^z \vec{A}_a |\lambda_{\rm max} \rangle}
    {\binom{L}{N}\, \lambda_{\rm max}^{L-N} 
    (\langle \lambda_{\rm max}| {^{t_p}}\!\vec{B}_{a'}^* \vec{B}_a |\lambda_{\rm max} \rangle)^{N} } + O(L^{-1}) \nonumber \\
    &= \lambda_{\rm max}^{-2} \langle \lambda_{\rm max}| {^{t_p}}\!\vec{A}_{a'}^* S^z \sigma^z_a \vec{A}_a |\lambda_{\rm max} \rangle 
    \langle \lambda_{\rm max}| {^{t_p}}\!\vec{A}_{a'}^* \sigma^z_a S^z \vec{A}_a |\lambda_{\rm max} \rangle +O(L^{-1}). 
\end{align}
\end{widetext}
While the contribution from the largest eigenvalue converges rapidly with increasing $L$, the system-size dependence arising from the combinatorial factors retains a nontrivial dependence on $N$. 
We note that the combinatorial hierarchy between contributions associated with different particle numbers is controlled by $N^2/L$ and remains well separated for $N^2 \ll L$; the fixed-$N$ limit considered here automatically satisfies this condition. 

We remark again that our evaluation of the string order parameter is valid only for fixed $N$, since, for states with mixed particle number, the subleading contribution from the $N$-particle sector becomes comparable to the leading contribution from the $(N-1)$-particle sector. As a result, the combinatorial hierarchy underlying the leading-order evaluation breaks down.  
For this reason, the method presented in this paper does not directly apply to states with mixed particle number, such as a coherent state $e^{a Q^{\pm}} |\Psi_0 \rangle$. Nevertheless, for such a coherent state, the string order parameter remains finite, reflecting the persistence of nontrivial topological structure, although its value differs from that of the ground state~\cite{bib:Katsura}.
%}

\bibliographystyle{apsrev4-2}
\nocite{apsrev42Control}
\bibliography{control,ref}

\end{document}